\documentclass[aps,prd,twocolumn,preprintnumbers,superscriptaddress,floatfix]{revtex4}

\usepackage{multirow, graphicx,amssymb,url,mathrsfs,amsmath}
\usepackage{eucal,wrapfig,boxedminipage,setspace,subfigure}
\usepackage{amsxtra,amstext,latexsym,dsfont}

\def\IR{{\hbox{{\rm I}\kern-.2em\hbox{\rm R}}}}
\def\IB{{\hbox{{\rm I}\kern-.2em\hbox{\rm B}}}}
\def\IN{{\hbox{{\rm I}\kern-.2em\hbox{\rm N}}}}
\def\IC{\,\,{\hbox{{\rm I}\kern-.59em\hbox{\bf C}}}}
\def\IZ{{\hbox{{\rm Z}\kern-.4em\hbox{\rm Z}}}}
\def\IP{{\hbox{{\rm I}\kern-.2em\hbox{\rm P}}}}
\def\IH{{\hbox{{\rm I}\kern-.4em\hbox{\rm H}}}}
\def\ID{{\hbox{{\rm I}\kern-.2em\hbox{\rm D}}}}

\def\det{{\rm det}}

\newcommand{\sgn}[1]{\mbox{sgn}(#1)}

\newcommand{\be}{\begin{equation}}
\newcommand{\ee}{\end{equation}}
\newcommand{\bea}{\begin{eqnarray}}
\newcommand{\eea}{\end{eqnarray}}

\newcommand{\mt}[1]{\textrm{\tiny #1}}

\def\rh {r_\mt{H}}

\begin{document}

\voffset 1cm

\newcommand\sect[1]{\emph{#1}---}

\title{Holographic complexity of anisotropic black branes}

\author{Seyed Ali Hosseini Mansoori}
\email{ shossein@ipm.ir, shosseini@shahroodut.ac.ir}
\affiliation{ Faculty of Physics, Shahrood University of Technology,
P.O.Box 3619995161 Shahrood, Iran}
\affiliation{ School of Astronomy, 
Institute for Research in Fundamental Sciences (IPM),
P.O. Box 19395-5531, Tehran, Iran}

\author{Viktor Jahnke}\email{ viktor.jahnke@correo.nucleares.unam.mx}
\affiliation{ Departamento de F\'isica de Altas Energias, Instituto de Ciencias Nucleares, Universidad Nacional Aut\'onoma de M\'exico Apartado Postal 70-543, CDMX 04510, M\'exico }

\author{Mohammad M. Qaemmaqami}
\email{m.qaemmaqami@ipm.ir}
\affiliation{ School of Particles and Accelerators
Institute for Research in Fundamental Sciences (IPM),
P.O. Box 19395-5531, Tehran, Iran}

\author{Yaithd D. Olivas
}
\email{yaithd.olivas@correo.nucleares.unam.mx}
\affiliation{ Departamento de F\'isica de Altas Energias, Instituto de Ciencias Nucleares, Universidad Nacional Aut\'onoma de M\'exico Apartado Postal 70-543, CDMX 04510, M\'exico }

\begin{abstract}

\noindent 
We use the complexity = action (CA) conjecture to study the full-time dependence of holographic complexity in anisotropic black branes. We find that the time behavior of holographic complexity of anisotropic systems shares a lot of similarities with the behavior observed in isotropic systems. In particular, the holographic complexity remains constant for some initial period, and then it starts to change so that the complexity growth rate violates the Lloyd's bound at initial times, and approaches this bound from above at later times. Compared with isotropic systems at the same temperature, the anisotropy reduces the initial period in which the complexity is constant and increases the rate of change of complexity. At late times the difference between the isotropic and anisotropic results is proportional to the pressure difference in the transverse and longitudinal directions. In the case of charged anisotropic black branes, we find that the inclusion of a Maxwell boundary term is necessary to have consistent results. Moreover, the resulting complexity growth rate does not saturate the Lloyd's bound at late times.
\end{abstract}

\maketitle

\newpage

\section{Introduction}\label{sec-intro}

The gauge-gravity duality \cite{duality1} provides a framework in which one can study the emergence of gravity from non-gravitational degrees of freedom. Within this framework, the gravitational theory lives in a higher dimensional space $\mathcal{M}$, usually called bulk, and the non-gravitational theory can be thought of as living on the boundary of $\mathcal{M}$. Despite the existence of a dictionary \cite{duality2,duality3} relating bulk and boundary quantities, the description of the black hole's interior in terms of boundary degrees of freedom remains elusive. Recently, there has been progress in this direction, with the conjecture that the growth of the interior of a black hole is related to the quantum computational complexity \footnote{The quantum computational complexity is a state-dependent quantity that measures how difficult is to prepare a given state. More precisely, one starts from some reference state and defines the complexity of the target state as the minimal number of simple unitary operations required to prepare it. For a more precise definition, see the review \cite{Aaronson-2016}.} of the states in the boundary theory. There are two main proposals relating the complexity to geometric quantities in the bulk, namely, the Complexity = Volume (CV) \cite{CV-1,CV-2} and the Complexity = Action (CA) \cite{CA-1,CA-2} conjectures. In the CV conjecture, the complexity is dual to the volume of a certain extremal surface in the bulk and provides an example of the recent proposal about the connection between tensor networks and geometry \cite{Swingle:2009bg,Vidal:2007hda,Hartman:2013qma}, while in the CA conjecture the complexity is dual to the gravitational action evaluated in certain region in the bulk. More details about CA conjecture will be given in section \ref{sec-HC}.

A convenient gravity set-up to study complexity growth is a two-sided black hole geometry. This geometry has two asymptotic regions, which we call left ($L$) and right ($R$) boundaries, and an Einstein-Rosen Bridge (ERB) connecting the two sides of the geometry. The Penrose diagram of this geometry is shown in figure \ref{fig-Penrose}. From the point of view of the boundary theory, the two-sided black hole is dual to a thermofield double (TFD) state, constructed out of two copies of the boundary theory \cite{eternalBH}
\begin{equation}
|TFD \rangle=\frac{1}{Z^{1/2}}\sum_n e^{\frac{-\beta E_n}{2}}e^{-i E_n (t_\mt{L}+t_\mt{R})} |E_n \rangle_\mt{L} | E_n \rangle_\mt{R}\,,
\end{equation}
where $L$ and $R$ label the quantum states of the left and right boundary theories, respectively. The TFD state is invariant under evolution with a Hamiltonian of the form $H=H_\mt{L}-H_\mt{R}$, which means that the system is invariant under the shifts  $t_\mt{L} \rightarrow t_\mt{L}+\Delta t $, and $t_\mt{R} \rightarrow t_\mt{R}-\Delta t $. As a result, the TFD state only depends on the sum of the left and right boundary times $t=t_\mt{L}+t_\mt{R}$. 

The ERB connecting the two sides of the geometry grows linearly with time. Classically, this behavior goes on forever. In \cite{CV-1} Susskind proposed that this behavior is dual to the growth of the computational complexity in the boundary theory, which is known to persist for very long times. Using the CV proposal, the authors of \cite{CV-2} showed that the late-time behavior of the rate of change of complexity is given by $d\mathcal{C}_V/dt=8\pi M/(d-1)$, where $M$ is the black hole's mass and $d$ is the number of dimensions of the boundary theory. 

Despite having a qualitative agreement with the behavior of complexity for quantum systems, the CV conjecture is defined in terms of an arbitrary length scale, which is usually taken to be of the order of the AdS radius. In order to avoid the ambiguity associated to the arbitrary length scale the authors of \cite{CA-1,CA-2} proposed the CA conjecture. For neutral black holes, the late time behavior of the rate of change of holographic complexity reaches a constant value which is also proportional to the black hole's mass 
\be
\lim_{t \rightarrow \infty}\frac{d\mathcal{C}_A}{dt} = \frac{2 M}{\hbar \pi}\,.
\ee
This late-time behavior may be associated with the Lloyd's bound on the rate of computation by a system with energy $M$ \cite{Lloyd-99}. This saturation of the complexification bound lead to the conjecture that the black holes are the fastest computers in nature \cite{CA-2}. It was later shown that a more precise definition of $\mathcal{C}_A$ requires the introduction of joint and boundary terms, which were not present in the calculation of \cite{CA-1,CA-2}. In particular, it was shown that the CA proposal also have an ambiguity related to the parametrization of null surfaces \cite{Lehner-2016}. Using the boundary and joint terms derived in \cite{Lehner-2016,Carmi-2016}, the authors of \cite{Carmi-2017,Kim-2017a} showed that these ambiguities do not affect the late time behavior 
of $d \mathcal{C}_A/dt$, but they play a role at early times, leading to a violation of Lloyd's bound.

Therefore, both the CA and the CV proposals have ambiguities which (apparently) cannot be eliminated. This is not a problem, however, because the same ambiguities were found in the definition of complexity for free quantum field theories \cite{Jefferson-2017,Chapman-2017,Khan-2018}. Moreover, the quantitative disagreement between the results obtained with the CA and CV proposals might be related to other ambiguities in the definition of complexity, like the choice of the reference state or the choice of the elementary gates. 

The Lloyd's bound was shown to be violated even at late times by anisotropic systems, including the SYM theory defined in a non-commutative geometry \cite{Couch-2017}, and Lifshitz and hyperscaling violating geometries \cite{Alishahiha-2018,Swingle-2017,An-2018}. This raises the question of whether there is a more general bound that is also respected by anisotropic systems. With this in mind, in this paper we use the CA conjecture to study the holographic complexity of a class of anisotropic black branes \footnote{Some previous work on holographic complexity include, for instance, \cite{Alishahiha-2015,Alishahiha-2017a,Alishahiha-2017b,Couch-2016,Chapman-2016,Cai-2016,Brown-2016,Brown-2017,Pan-2016,Yang-2016,Momeni:2016ekm,Momeni:2016ira,Kim-2017,Cai-2017,Bakhshaei-2017,Abad-2017,Tao-2017,Guo-2017,Nagasaki-2017,Miao-2017,Qaemmaqami-2017,
Sebastiani-2017,Cottrell-2017,Moosa-2017,HM-2017,Ge-2017,Zhang-2017,Moosa-2017a,Ovgun-2018,Agon-2018,Auzzi:2018zdu,An:2018dbz,Bamba:2018ouz,Ghaffarnejad:2018prc,Ghaffarnejad:2018bsd,Auzzi:2018pbc,Fareghbal:2018ngr,Nagasaki:2018csh}.}. 
More specifically, we consider the Mateos and Trancanelli (MT) model \cite{MT1,MT2}, the D'Hoker and Kraus (DK) model \cite{DHoker:2009mmn}, and the Cheng-Ge-Sin (CGS) model  \cite{Cheng:2014qia,Cheng:2014sxa}, and study the time dependence of holographic complexity in thermofield double states which are dual to two-sided black brane geometries.  

The MT model is a solution of type IIB supergravity that was designed to model the effects of anisotropy in the quark-gluon plasma (QGP) created in heavy ion collisions. The anisotropy is present in the initial stages after the collision and it leads to different transverse and longitudinal pressures in the plasma. For our purposes, the main motivation to consider this model is that it describes a renormalization group (RG) flow from an AdS geometry in the ultraviolet (UV) to a Lifshitz-like geometry in the infrared (IR). The transition is controlled by the ratio $a/T$, where $a$ is a parameter that measures the degree of anisotropy and $T$ is the black brane's temperature. From the point of view of the boundary theory, this parameter is small close to the UV fixed point and large close the IR fixed point. We would like to understand how the complexity rate changes as we move along this RG flow and whether this system respects the Lloyd's bound. As a first step towards this, we have considered small deviations from the UV fixed point, i.e., small values of $a/T$, which can be incorporated by considering an analytical black brane solution with small corrections due to anisotropy \footnote{To investigate the entire RG flow it is necessary to consider arbitrary values of $a/T$. This leads to technical difficulties because the solution needs to be calculated numerically in these cases. We are currently investigating these more general cases.}. 

The DK model is a solution of 5-dimensional Einstein-Maxwell gravity that is dual to the 4-dimensional $\mathcal{N}=4$ SYM theory in the presence of a background magnetic field. This solution describes an RG flow between an AdS geometry in the UV to a BTZ $\times \mathbb{R}^2$ geometry in the IR. The parameter controlling such transition is $B/T^2$, where $B$ is the intensity of the magnetic field, while $T$ is the black brane's temperature.  We would like to understand how the magnetic field affects the rate of change of complexity.

The CGS model \cite{Cheng:2014qia,Cheng:2014sxa} is a generalization of the MT model to the charged case. 
The geometry is an anisotropic RN-AdS solution, being also affected by a charge parameter, $q$. When  $a\neq 0$ and $q \rightarrow 0$, the solution reduces to the MT geometry. When $q \neq 0$ and $a \rightarrow 0$, the solution becomes an RN-AdS geometry. The motivation to consider this solution is to understand how the rate of change of complexity is affected by the presence of uncharged and charged matter fields, and whether the CA prescription can provide sensible results in the presence of several matter fields.

In the uncharged cases, we find that the time behavior of holographic complexity is qualitatively similar to the behavior observed for isotropic systems, namely, the holographic complexity remains constant for some period, and then it starts to change so that the rate of complexity growth violates the Lloyd's bound at initial times, and it approaches this bound from above at later times.
Additionally, we find that the net effect of anisotropy is basically a vertical upward shift in the curves of the rate of change of holographic complexity versus time. At later times, the difference between the isotropic and anisotropic results is proportional to the difference in pressures in the longitudinal and transverse directions. In the charged case, we find that the inclusion of a Maxwell boundary term is necessary to have consistent results.


The remainder of paper is organized as follows. In section \ref{sec-grav} we review the MT and DK solutions and present some of its thermodynamic properties. In section \ref{sec-HC} we use the CA conjecture to study the full-time behavior of holographic complexity of thermofield double states which are dual to two-sided anisotropic black branes solutions. The case of charged anisotropic black branes is considered in section \ref{sec-charged}. We discuss our results in section \ref{sec-disc}. We relegate some technical details of the calculations to the appendices \ref{appA} and \ref{appB}.
\section{Gravity set-up}\label{sec-grav}
\subsection*{Anisotropic black branes: the MT model}
The Mateos and Trancanelli (MT) model \cite{MT1,MT2} is a solution of type IIB supergravity whose effective action in five dimensions can be written as
\begin{eqnarray}
S&=&\frac{1}{16\pi G_\mt{N}}  \int_{\mathcal{M}}  d^{5}x\sqrt{-g}\Big[R+\frac{12}{L^2}-\frac{1}{2}\left( \partial\phi\right)^2 \nonumber \\
 &-&\frac{1}{2}e^{2\phi}\left( \partial\chi \right)^2\Big]+S_\textrm{GH}\,,
\end{eqnarray}
where $\phi$, $\chi$ and $ g_{\mu\nu} $ are the dilaton field, the axion field and the metric respectively, $G_\mt{N}$ is the five-dimensional Newton constant, and $S_\textrm{GH}$ is the Gibbons-Hawking term. The solution in Einstein frame takes the form
\begin{eqnarray}
ds^2&=&L^2 \,e^{-\phi(r)/2}\Big[ -r^2 \mathcal{F}(r)\, \mathcal{B}(r)\, dt^2 \nonumber \\
 &+& \frac{dr^2}{r^2 \mathcal{F}(r)}+r^2 \left(dx^2+dy^2+\mathcal{H}(r)\,dz^2 \right)\Big],
\label{eq-metric}
\end{eqnarray}
with
\begin{equation} 
\chi = a \,z\,,\,\,\,\, \phi = \phi(r)\,,\,\,\,\,\mathcal{H}=e^{-\phi}\,,
\label{eq-axionDilaton}
\end{equation}
where $(t,x,y,z)$ are the gauge theory coordinates and $r$ is the AdS radial coordinate. Here $L$ is the AdS radius, which we set to unity in the following \footnote{Note that $L=1$ implies that $G_\mt{N}=\frac{\pi}{2N^2}$ (see e.g. \cite{HotQCD}), where $N$ is the rank of the gauge group $SU(N)$ of the dual field theory.}. The above solution has a horizon at $r=\rh
$ and the boundary is located at $r = \infty$, where $\mathcal{F}=\mathcal{B}=\mathcal{H}=1$ and $\mathcal{\phi}=0$. The axion is proportional to the $z-$coordinate and this introduces an anisotropy into the system, which is measured by the anisotropy parameter $a$. For $a \neq 0$, the above solution corresponds to the gravity dual of $\mathcal{N}=4$ SYM theory, with gauge group $SU(N)$, deformed by a position-dependent theta term. When $a=0$, the above solution reduces to the gravity dual of the undeformed SYM theory. The functions $\mathcal{F}$, $\mathcal{B}$, $\mathcal{H}$ and the dilaton $\mathcal{\phi}$ can be determined analytically \footnote{For generic values of the anisotropy parameter, the metric functions can be determined numerically. For more details, see the appendix A of \cite{MT2}.} for small values of the anisotropy parameter $a$ as
\begin{eqnarray}
\mathcal{F}&=&1-\frac{\rh^4}{r^4}+\frac{a^2}{24 r^4 \rh^2} \Big[8r^2\rh^2-2r_H^2(4+5\log 2)\nonumber \\
&+&(3r^4+7\rh^4)\log \left(1+\frac{\rh^2}{r^2} \right) \Big]+\mathcal{O}(a^4)\\
\mathcal{B}&=&1-\frac{a^2}{24 \rh^2} \left[\frac{10 \rh^2}{r^2+\rh^2}+\log \left(1+\frac{\rh^2}{r^2} \right) \right]+\mathcal{O}(a^4) \\
\phi&=&-\frac{a^2}{4\rh^2} \log \left(1+\frac{\rh^2}{r^2} \right)+\mathcal{O}(a^4)\,.
\end{eqnarray}
By requiring regularity of the Euclidean continuation of the above metric at the horizon,
one obtains the Hawking temperature as
\begin{equation}
T=\frac{\rh}{\pi}+\frac{(5 \log 2-2)}{48\pi}\frac{a^2}{\rh}+\mathcal{O}(a^4)\,.
\label{eq-Trh}
\end{equation}
The Bekenstein-Hawking entropy can be obtained from the horizon area as
\begin{equation}
S =\frac{\rh^3}{4G_\mt{N}} \left(1+\frac{5a^2}{16 \rh^2} \right) V_3+\mathcal{O}(a^4)\,.
\label{eq-Srh}
\end{equation}
where $V_3 =\int dx dy dz$ is the volume in the $xyz-$directions. Using holographic renormalization, the stress tensor of the deformed SYM theory can be obtained as \cite{MT2,Jahnke-2014}
\be
T_{ij}=\text{diag}(E,P_{xy},P_{xy},P_{z})\,,
\ee
where
\bea
E=\frac{3\pi^2 N^2 T^4}{8}+\frac{N^2T^2}{32}a^2+\mathcal{O}(a^4)\,,
\eea

is the energy density of the black brane and
\begin{eqnarray}
P_{xy}&=&\frac{\pi^2 N^2 T^4}{8}+\frac{N^2T^2}{32}a^2+\mathcal{O}(a^4)\,,\\
 P_z&=& \frac{\pi^2 N^2 T^4}{8}-\frac{N^2T^2}{32}a^2+\mathcal{O}(a^4)\,
\end{eqnarray}
are the pressures along the transverse and longitudinal directions, respectively.
The mass of the black brane can then be calculated as
\be
M=E\,V_3 = \left( \frac{3\pi^2 N^2 T^4}{8}+\frac{N^2T^2}{32}a^2 \right) V_3+\mathcal{O}(a^4)\,,
\label{eq-M}
\ee
A more simple way of calculating the black brane's mass is through the expression
\begin{eqnarray}
\nonumber  &&M=\int T\,dS=\int_{0}^{\rh}T(\rh)\frac{dS(\rh)}{d \rh}d\rh \\
&& =\frac{V_3}{16\pi G_\mt{N}}\Big[3\rh^4
 +\frac{\rh^2 a^2}{4}\left(5 \log2-1 \right) \Big]+\mathcal{O}(a^4)
\label{eq-mass}
\end{eqnarray}
where the integral was calculated using the equations (\ref{eq-Trh}) and (\ref{eq-Srh}) for $T(\rh)$ and $S(\rh)$, respectively. Expressing $\rh$ as a function of the temperature $T$ and using that $G_\mt{N} =\pi/(2N^2)$, we recover the expression for the mass given in equation (\ref{eq-M}).

Note that the mass of the anisotropic black brane is larger than the mass of an isotropic black brane with the same temperature, or with the same horizon radius. For future reference, we note that
\be
M(a)=M(0)+\frac{V_3}{2}\left( P_{xy}-P_z\right)+\mathcal{O}(a^4)\,.
\label{eq-compMass}
\ee

\subsection*{Magnetic black branes: the DK model}

The D'Hoker and Kraus (DK) model \cite{DHoker:2009mmn} is a magnetic black brane solution of 5-dimensional Einstein-Maxwell gravity. The action of this model reads
\be
S=\frac{1}{16\pi G_\mt{N}} \int d^5x \sqrt{-g} \left(R+12 -F_{MN}F^{MN} \right)\,.
\ee
For very large values of the magnetic field ($B/T^2>>1$), the solution takes the form \footnote{For small and intermediate values of $B/T^2$, the geometry can be found numerically. See \cite{DHoker:2009mmn} for more details.}
\bea
ds^2=-3(r^2-\rh^2)dt^2+\frac{dr^2}{3(r^2-\rh^2)} \nonumber\\
-\frac{B}{\sqrt{3}}\left(dx^2+dy^2 \right)+3r^2 dz^2\,.
\label{eq-solB}
\eea
with field strength $F=B \,dx \wedge dy$. 

The Hawking temperature and the Bekenstein-Hawking entropy associated to the above solution are easily found to be
\be
T=\frac{3 \rh}{2 \pi}\,,\,\,\,\,\,\,S=\frac{3 V_3 B \, \rh^2}{4 G_\mt{N}}\,.
\label{T-DK}
\ee
where $V_3=\int dx dy dz$. The black brane's mass can then be calculated as
\be
M_B=\int T dS =\frac{V_3}{16 \pi G_\mt{N}} \times 3B \rh^2\,.
\ee

\subsection*{Penrose diagram}
\begin{figure}
\centering
 \includegraphics[width=8.5cm]{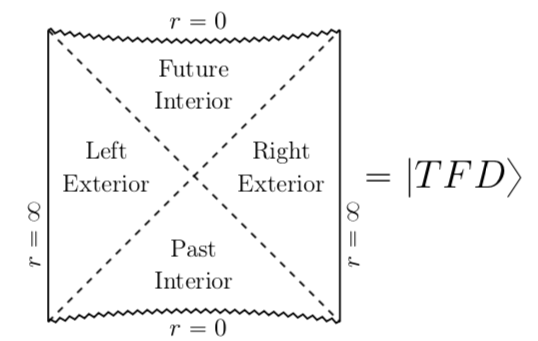} 
\caption{ Penrose diagram for the two-sided black branes we consider. This geometry is dual to a thermofield double state constructed out of two copies of the boundary theory.}
\label{fig-Penrose}
\end{figure}

Lastly, we comment that the above gravitationals solution can be extended to a two-sided eternal black brane geometry, with two asymptotic boundaries. See figure \ref{fig-Penrose}. The extended solution is dual to a thermofield double state constructed out of two copies of the boundary theory.  

The Penrose diagram is obtained as follows. We consider a general general metric of the form given in Eq. (\ref{eq-metricGen}). We first define Kruskal-Szekeres coordinates $U$ and $V$ as
\begin{equation}
\begin{split}
U&=+e^{\frac{2\pi}{\beta}\left(r_*-t\right)}\,,\,\,V=-e^{\frac{2\pi}{\beta}\left(r_*+t\right)} \,\,\,(\text{left exterior region})  \\
U&=-e^{\frac{2\pi}{\beta}\left(r_*-t\right)}\,,\,\,V=+e^{\frac{2\pi}{\beta}\left(r_*+t\right)} \,\,\,(\text{right exterior region})  \\
U&=+e^{\frac{2\pi}{\beta}\left(r_*-t\right)}\,,\,\,V=+e^{\frac{2\pi}{\beta}\left(r_*+t\right)} \,\,\,(\text{future interior region}) \\
U&=-e^{\frac{2\pi}{\beta}\left(r_*-t\right)}\,,\,\,V=-e^{\frac{2\pi}{\beta}\left(r_*+t\right)} \,\,\,(\text{past interior region}) 
\end{split}
\end{equation}
where $\beta$ is the black hole inverse temperature, and $r_*$ is the tortoise coordinate, which is defined in (\ref{eq-tortoise}). In terms of these coordinates, the metric (\ref{eq-metricGen}) becomes
\be
ds^2= -\frac{\beta^2e^{-\frac{4\pi}{\beta}r_*}}{8\pi^2}  G_{tt}(UV) dU dV+G_{ij}(UV) dx^i dx^j\,.
\ee
The Penrose diagram is obtained with one additional change of coordinates, $\tilde{U} =\tan^{-1} (U)$ and $\tilde{V} =\tan^{-1} (V)$, in terms of which the boundaries of the spacetime lie at finite coordinate distance. The Penrose diagram will have the form given in figure \ref{fig-Penrose} as long as the blackening factor $\mathcal{F}(r)$ has a single root, and the tortoise satisfies three conditions, namely: (I) $\lim_{r \rightarrow \infty} r_*(r) =0$; (II) $\lim_{r \rightarrow \rh} r_*(r) =-\infty$; (III) $\lim_{r \rightarrow 0} r_*(r) =0$. Each point in the Penrose diagram is a three-dimensional space, with metric $G_{ij}$. The fact that $G_{ij}$ is anisotropic does not affect the diagram, because the diagram is only constructed out of the coordinates $t$ and $r_*$. We explicitly checked that both the MT and the DK models satisfy the above conditions. The Penrose diagram of the charged MT model, considered in section \ref{sec-charged}, is different because in that case the blackening factor $\mathcal{F}(r)$ has two roots.

\section{Holographic Complexity}\label{sec-HC}
In this section we compute the holographic complexity using the complexity=action (CA) \cite{CA-1,CA-2}. Here we follow closely the analysis of \cite{Carmi-2017}, with adaptations for anisotropic systems. We consider neutral anisotropic black branes with a generic bulk action of the form
\begin{equation}
S=\frac{1}{16\pi G_\mt{N}}\int d^dx dr\, \sqrt{-g} \mathcal{L}(r,x)\,,
\label{eq-Sgen}
\end{equation}
and metric
\begin{equation}
ds^{2}=-G_{tt}(r) dt^2 +G_{rr}(r) dr^2 +G_{ij}(r) dx^{i}dx^{j}
\label{eq-metricGen}
\end{equation}
where $r$ is the AdS radial coordinate and $(t,x^i)$ are the gauge theory coordinates. Here $i=1,2,...,d-1$. We take the boundary as located at $r = \infty$ and we assume the existence of a horizon at $r=\rh$, where $G_{tt}$ has a zero and $G_{rr}$ has a  simple pole. We denote as $G$ the determinant of $G_{ij}$, i.e. $G=\det(G_{ij})$.

In the computations of holographic complexity it is convenient to use coordinates that cover smoothly the two sides of the geometry. We use Eddington-Finkelstein coordinates
\begin{equation}
u=t-r^*(r)\,,\,\,\,\,v=t+r^*(r)\,,
\end{equation}
where the tortoise coordinate is defined as
\begin{equation}
r^*(r)=\textrm{sgn}(G_{tt}(r)) \int^r dr' \sqrt{\frac{G_{rr}(r')}{G_{tt}(r')}}\,.
\label{eq-tortoise}
\end{equation}

\begin{figure}
\centering
{\includegraphics[width=7.5cm]{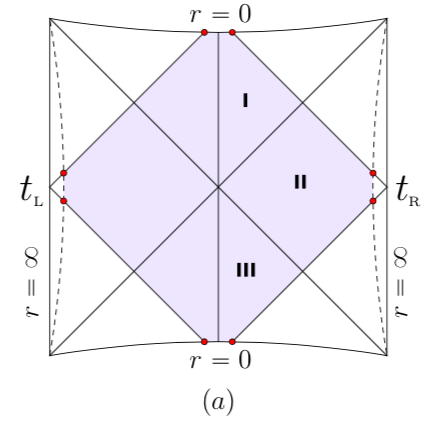}
\includegraphics[width=7.5cm]{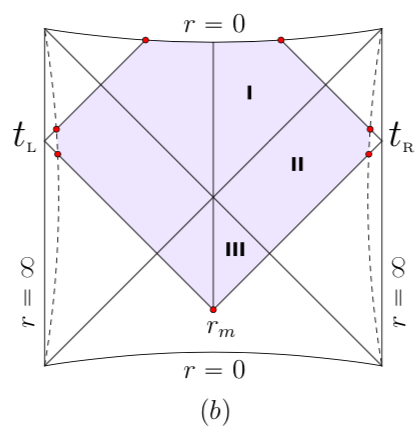}}
\caption{ Penrose diagram and the WDW patch (blue region) for the two-sided black brane we consider. (a) Configuration at initial times ($t \leq t_c$) in which the WDW patch intersects both the future and the past singularity. (b) Configuration at later times ($t > t_c$) when the WDW patch no longer intersects the past singularity. The dashed lines represent the cutoff surfaces at $r=r_\mt{max}$.}
\label{fig-WDW}
\end{figure}

The CA conjecture states that the quantum complexity of the state of the boundary theory is given by the gravitational action evaluated in a region of the bulk known as the Wheeler-DeWitt (WDW) patch
\begin{equation}
\mathcal{C}_A=\frac{I_\mt{WDW}}{\pi \hbar}\,.
\end{equation}
The WDW patch is the domain of dependence of any spatial slice anchored at a given pair of boundary times $(t_\mt{L},t_\mt{R})$. See figure \ref{fig-WDW}. The gravitational action in the WDW patch is divergent because this region extends all the way up to the asymptotic boundaries of the space-time. We regularize this divergence by introducing a cutoff surface at $r=r_\mt{max}$ near the boundaries. We also introduce a cutoff surface $r=\epsilon_0$ near to the past and future singularities.
Without loss of generality, we consider the time evolution of holographic complexity for the symmetric configuration $t_\mt{L}=t_\mt{R}=t/2$. More general cases can be obtained from the symmetric configuration by using the fact that the system is symmetric under shifts $t_\mt{L} \rightarrow t_\mt{L}+\Delta t$ and $t_\mt{R} \rightarrow t_\mt{R}-\Delta t$.
The gravitational action in the WDW patch can be written as
\begin{equation}
I_{\textrm{WDW}}=I_{\textrm{bulk}}+I_{\textrm{surface}}+I_{\textrm{joint}}\,,
\end{equation}
where
\begin{equation}
I_{\textrm{bulk}}=\frac{1}{16\pi G_\mt{N}}  \int_{\mathcal{M}}  d^{d+1}x\sqrt{-g} \mathcal{L}(x)
\end{equation}
is the bulk action and $I_{\textrm{surface}}$ and $I_{\textrm{joint}}$ are surface and joint terms that are necessary to have a well-defined variational principle when one considers a finite domain of space-time \cite{Lehner-2016}. The surface terms are given by
\begin{equation}
I_{\textrm{surface}}=\frac{1}{8\pi G_\mt{N}}\int_{\mathcal{B}}d^{d}x\sqrt{|h|}K \pm \frac{1}{8\pi G_{N}}\int_{\mathcal{B}'}d\lambda d^{d-1}\theta\sqrt{\gamma}\kappa
\end{equation}
where the first term, which is defined in terms of the trace of the extrinsic curvature $K$, is the well-known Gibbons-Hawking-York boundary term \cite{York-72,GH-77}. This term is necessary when the boundary includes (smooth) space-like and time-like segments, which we denoted as $\mathcal{B}$. The second term in the above equation includes the contribution of null segments. This term is defined in terms of the parameter $\kappa$, which measure how much the null surface $\mathcal{B}'$ fails to be affinely parametrized. Here we follow \cite{Lehner-2016} and set $\kappa=0$, so that we do not need to consider these null boundary terms. This choice of $\kappa$ correspond to affinely parametrize the null boundary surfaces.

The joint terms are necessary when the intersection of two boundary terms is not smooth. These terms can be written as
\begin{equation}
I_{\textrm{joint}}=\frac{1}{8\pi G_\mt{N}}  \int_{\Sigma}d^{d-1}x\sqrt{\sigma}\eta + \frac{1}{8\pi G_{N}}\int_{\Sigma'}d^{d-1}x\sqrt{\sigma}\bar{a}
\end{equation}
where the first term \footnote{This contribution is known as the Hayward joint term \cite{Hayward-93,Brill-94}.} corresponds to the intersection of two boundary segments which can be time-like or space-like, so the intersection can be of the type: time-like/time-like, time-like/space-like or space-like/space-like. As the WDW patch do not include such intersection, we do not need to consider this first term. The second term includes the contribution of the intersection of a null segment with any other boundary segment, so it includes contribution of the type: null/null, null/time-like and null/space-like. A more precise definition of the surface and joint terms will be given throughout the text along with the adopted conventions \footnote{Here we adopted the conventions found in the appendix A of \cite{Carmi-2016}.}. The quantity $\bar{a}$ is defined in appendix \ref{appA}.

\begin{figure}
\centering
\includegraphics[width=9cm]{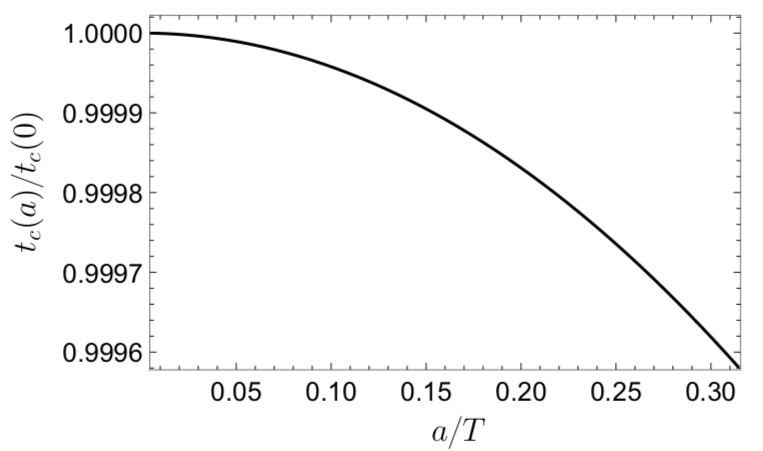} 
\caption{
Critical time (normalized by isotropic result) versus $a/T$. We consider increasing values of $a$, but we choose $\rh$ in such a way to keep fixed the temperature as $T=1/\pi$.}
\label{fig-tc}
\end{figure}

As first pointed out in \cite{Chapman-2017}, at early times the WDW patch intersects both the future and the past singularity, and this causes $I_{\textrm{WDW}}$ to be constant for some period of time $0 \leq t \ \leq t_c$. At later times, $t > t_c$, the WDW patch no longer intersects the past singularity, and $I_{\textrm{WDW}}$ starts to change with time. These two cases are illustrated in figure \ref{fig-WDW}. The time scales separating these two regimes can be written as
\begin{equation}
t_c =2\left(r^*_{\infty}- r^*(0)\right)\,,\,\,\,\,\,\, r^*_{\infty}=\lim_{r \rightarrow \infty} r^*(r)
\label{eq-tc}
\end{equation}
where we have used that $t_\mt{L}=t_\mt{R}=t/2$. 
Figure \ref{fig-tc} shows how the critical time (\ref{eq-tc}) behaves as a function of the anisotropy parameter in MT model. This figure shows that, as compared to an isotropic system at the same temperature, the anisotropy reduces the critical time, i.e., the complexity starts to change earlier in anisotropic systems.
\subsection{Behavior at initial times: $0 \leq t \leq t_c$}
For initial times $0 \leq t \leq t_c$ the WDW patch intersects with both the future and past singularities. The contributions for $I_{\textrm{WDW}}$ include: the bulk term, the GHY terms and the joint terms. In principle, the GHY terms include contributions from the cutoff surfaces at $r=r_{\textrm{max}}$ and $r=\epsilon_0$, as well as from the null boundaries of the WDW patch. However, since we affinely parametrize the null surfaces, we do not need to consider the surface contributions from the null boundaries. The joint terms include contribution from the intersection of the null boundaries of the WDW patch with the cutoff surfaces at $r=r_{\textrm{max}}$ and $r=\epsilon_0$. We use the left-right symmetry of the WDW patch to calculate $I_{\textrm{WDW}}$ for the right side of Penrose diagram and then multiply the result by two.

To calculate the bulk contributions we split the right-side of the WDW patch into three parts: region I, region II and region III, which are shown in figure \ref{fig-WDW} (a). We then calculate the bulk contribution as
\begin{equation}
I_{\textrm{bulk}}(t\leq t_c)=2\left( I_{\mt{bulk}}^\mt{I}+I_{\mt{bulk}}^\mt{II}+I_{\mt{bulk}}^\mt{III} \right)\,,
\end{equation}
where \footnote{Here we are using that $I_\textrm{bulk}^{I}=\frac{1}{16\pi G_\mt{N}}  \int_{I}  d^{d+1}x\sqrt{-g} \, \mathcal{L}(x)=\frac{1}{16\pi G_\mt{N}}  \int_{I}  d^{d-1}x \int dr \sqrt{G_{tt}G_{rr}G} \, \mathcal{L}(r) \int_{0}^{v_1-r^*(r)} dt$, where $v_1=t_R+r^{*}_{\infty}$ defines the future null boundary of the right side of the WDW patch. In region $II$, for instance, the integral over the time coordinate is $\int_{u_1+r^*(r)}^{v_1-r^*(r)} dt=2\Big( r^{*}_{\infty}-r^{*}(r)\Big)$, where $u_1=t_R-r^{*}_{\infty}$ defines the past null boundary of the right side of the WDW patch.}
\bea
\nonumber I_\mt{bulk}^\mt{I}&=&\frac{V_{d-1}}{16\pi G_\mt{N}} \int_{\epsilon_{0}}^{\rh} dr \sqrt{-g}\, \mathcal{L}(r) \left(\frac{t}{2}+r^{*}_{\infty}-r^{*}(r) \right)\\
I_\mt{bulk}^\mt{II}&=&\frac{V_{d-1}}{8\pi G_\mt{N}} \int_{\rh}^{r_\mt{max}} dr \sqrt{-g} \, \mathcal{L}(r)\Big( r^{*}_{\infty}-r^{*}(r) \Big)\nonumber \\
\nonumber I_\mt{bulk}^\mt{III}&=&\frac{V_{d-1}}{16\pi G_\mt{N}} \int_{\epsilon_{0}}^{\rh} dr \sqrt{-g} \, \mathcal{L}(r)\left(-\frac{t}{2}+r^{*}_{\infty}-r^{*}(r) \right)\\
\eea
with  $V_{d-1}=\int d^{d-1}x$. Note that in the above expressions we are assuming that the on-shell Lagrangian $\mathcal{L}$ only depends on $r$. 
Summing  all the contributions we obtain 
\be
I_\mt{bulk}=\frac{1}{2\pi G_\mt{N}} \int_{\epsilon_{0}}^{r_{\textrm{max}}} dr \sqrt{-g} \,\mathcal{L}(r)\Big( r^{*}_{\infty}-r^{*}(r) \Big)\,.
\label{eq-Ibulk}
\ee
Note that $I_\textrm{bulk}(t\leq t_c)$ does not depend on time.
Now we turn to the computation of the GHY surface terms. These contribution come from the cutoff surfaces at $r=r_{\textrm{max}}$ on the two sides of the geometry and from the cutoff surfaces at $r=\epsilon_0$ both at the past and future singularities. In either cases the surfaces are described by a relation of the form $r=\text{constant}$, and the outward-directed normal vector are proportional to $\partial_{\mu} (r-\text{constant})$. We write the corresponding normal as
\be
n_{\mu}=(n_t,n_r,n_i)= b\,(0,1,0)\,,
\ee
where $b$ is some normalization constant. We normalize the normal vector as $n^2=n^r n_r=\pm1$, where the plus sign is for space-like vectors at the $r=r_\mt{max}$ cutoff surface, and the minus sign if for the time-like vectors at the $r=\epsilon_0$ cutoff surface. We obtain
\bea
 n^{(s)}_{\mu}&=&(n_t^{(s)},n_r^{(s)},n_i^{(s)})=(0,\sqrt{G_{rr}(r_\mt{max})},0)\\
 n^{(t)}_{\mu}&=&(n_t^{(t)},n_r^{(t)},n_i^{(t)})=(0,\sqrt{-G_{rr}(\epsilon_0)},0).
\eea
where the superscript $(s)$ denotes space-like vectors, while the superscript $(t)$ denotes time-like vectors. The trace of the extrinsic curvature of these $r-$constant surfaces can be calculated as
\bea
K&=\nabla_{\mu}n^{\mu}=\frac{1}{\sqrt{-g}}\partial_{r}\left(\sqrt{-g} n^{r}\right)\Big|_{r=\epsilon_{0}, r_{\mt{max}}} \nonumber\\
&=\frac{ 1}{2\sqrt{\mp G_{rr}}}\left[\frac{\partial_{r}G_{tt}}{G_{tt}}+\frac{\partial_{r}G}{G} \right]\Big|_{r=\epsilon_{0}, r_{\mt{max}}}
\label{eq-K}
\eea
where we use the minus sign for the $r=\epsilon_0$ surface and the plus sign for the $r=r_\mt{max}$ surface. Here $G=\det (G_{ij})$ is the determinant along the transverse coordinates $x^i$, not the full determinant, which we denoted as $g$.

The GHY surface contributions can then be written as
\be
I_\mt{surface}(t\leq t_c)= I_{\mt{surface}}^{\mt{future}}+I_{\mt{surface}}^{\mt{past}}+I_{\mt{surface}}^{\mt{bdry}}
\ee
where the contributions from the cutoff surfaces at future and past singularities are given by
\bea
I_{\mt{surface}}^{\mt{future}}&=&\frac{V_{d-1}}{8\pi G_\mt{N}} \mathcal{G}(r) \left( \frac{t}{2}+r^{*}_{\infty}-r^{*}(r) \right)\Big|_{r=\epsilon_{0}} \nonumber\\
I_{\mt{surface}}^{\mt{past}}&=&\frac{V_{d-1}}{8\pi G_\mt{N}} \mathcal{G}(r)\left( -\frac{t}{2}+r^{*}_{\infty}-r^{*}(r) \right)\Big|_{r=\epsilon_{0}} 
\eea
and the contributions from the cutoff surfaces at the two asymptotic boundaries read
\be
I_{\mt{surface}}^{\mt{bdry}}=\frac{V_{d-1}}{8\pi G_\mt{N}} \mathcal{G}(r) \Big(r^{*}_{\infty}-r^{*}(r) \Big)\Big|_{r=r_{\mt{max}}}\
\ee
where 
\be
\mathcal{G}(r)=\sqrt{\frac{G_{tt}G}{G_{rr}}}\left[\frac{G_{tt}'}{G_{tt}}+\frac{G'}{G} \right]\,.
\ee
In the above expressions we have already multiplied the results by two to account for the two sides of the WDW patch. Note that $I_{\mt{surface}}^{\mt{bdry}}$ does not depend on time. Moreover, the time dependence of $I_{\mt{surface}}^{\mt{future}}$ and $I_{\mt{surface}}^{\mt{past}}$ cancel, so that the total surface contribution is time-independent
\bea
 I_\mt{surface}(t\leq t_c)&=& \frac{V_{d-1}}{4\pi G_\mt{N}} \mathcal{G}(r)\left( r^{*}_{\infty}-r^{*}(r) \right)\Big|_{r=\epsilon_{0}}\nonumber \\  &+&\frac{V_{d-1}}{8\pi G_\mt{N}} \mathcal{G}(r) \left( r^{*}_{\infty}-r^{*}(r) \right)\Big|_{r=r_\mt{max}}
\label{eq-Isurface}
\eea
The only terms left to calculate are the joint contributions that come from the intersections of the null boundaries of the WDW patch with the cutoff surfaces at $r=r_\mt{max}$ and $r=\epsilon_0$. 
The joint terms can be written as
\be
I_\mt{joint}=I_\mt{joint}^\mt{sing}+I_\mt{joint}^\mt{bdry}\,,
\ee
where $I_\mt{joint}^\mt{sing}$ includes the contributions from the past and future singularities and $I_\mt{joint}^\mt{bdry}$ corresponds to the contribution from the two asymptotic boundaries. In \cite{Chapman-2016} it was shown that, for a large class of isotropic systems, the contribution from the asymptotic boundaries do not depend on time, while the contributions at $r=\epsilon_0$ vanish. We show in appendix \ref{appA} that this also happens in anisotropic systems. So we can write
\be
I_\mt{joint}(t\leq t_c)=I_\mt{joint}^\mt{bdry}\,,
\ee
where $I_\mt{joint}^\mt{bdry}$ does not depend on time.

Finally, as none of the terms $I_\mt{bulk}$,  $I_\mt{surface}$ and  $I_\mt{joint}$ depend on time for $0 \leq t \leq t_c$, the gravitational action evaluated on the WDW patch is constant for this period of time
\be
\frac{dI_\mt{WDW}}{dt}=0\,,\,\,\,\,\,\,\text{for}\,\,0\leq t \leq t_c\,.
\ee
\subsection{Behavior at later times: $t > t_c$}
For later times $t > t_c$ the WDW patch no longer intersects with the past singularity. In this case, there are no surface and joint terms related to the past singularity, but there is an additional joint term that comes from the intersection of two null boundaries of the WDW patch. See figure \ref{fig-WDW} (b). Again, we calculate all the contribution for the right side of the WDW patch and multiply the results by two to account for the two sides of the geometry.

To compute the bulk contribution, we again split the right side of the WDW patch into three regions, which we call I, II and III. See figure \ref{fig-WDW} (b). We write the total bulk contribution as
\be
I_{\mt{bulk}}(t > t_c)=2\left( I_{\mt{bulk}}^\mt{I}+I_{\mt{bulk}}^\mt{II}+I_{\mt{bulk}}^\mt{III} \right)\,,
\ee
where now
\bea
\nonumber I_\mt{bulk}^\mt{I}&=&\frac{V_{d-1}}{16\pi G_\mt{N}} \int_{\epsilon_{0}}^{\rh} dr \sqrt{-g} \, \mathcal{L}(r) \left(\frac{t}{2}+r^{*}_{\infty}-r^{*}(r) \right)\\
I_\mt{bulk}^\mt{II}&=&\frac{V_{d-1}}{8\pi G_\mt{N}} \int_{\rh}^{r_\mt{max}} dr \sqrt{-g} \, \mathcal{L}(r)\Big( r^{*}_{\infty}-r^{*}(r) \Big)\nonumber \\\nonumber 
I_\mt{bulk}^\mt{III}&=&\frac{V_{d-1}}{16\pi G_\mt{N}} \int_{r_m}^{\rh} dr \sqrt{-g}\, \mathcal{L}(r)\left(-\frac{t}{2}+r^{*}_{\infty}-r^{*}(r) \right)\\
\eea
where the only difference from the $0 \leq t \leq t_c$ case is that the $r-$integral in the region III starts at the point $r=r_m$, instead of starting at the cutoff surface $r=\epsilon_0$ at the past singularity. The point $r_m$  determines the intersection of the two past null boundaries of the WDW patch and it satisfies the equation
\be
\frac{t}{2}-r^{*}_{\infty}+r^{*}(r_m)=0\,.
\label{eq-rm}
\ee
which can be solved numerically. Note that  we recover the equation that gives the critical time $t_c$ when we take the limit $r_m \rightarrow 0$ in the above equation.
Summing the above contributions we can write the bulk term at later times as the bulk term at initial times plus a time-dependent term
\begin{eqnarray}
I_{\mt{bulk}}(t &>& t_c)=I_{\mt{bulk}}(t \leq t_c)+\nonumber\\
&& \nonumber \frac{V_{d-1}}{8\pi G_\mt{N}} \int_{\epsilon_{0}}^{r_{m}} dr  \sqrt{-g}\,\mathcal{L}(r)\left( \frac{t}{2}-r^{*}_{\infty}+r^{*}(r) \right)\,\\
\end{eqnarray}
where $I_{\mt{bulk}}(t \leq t_c)$ is given in equation (\ref{eq-Ibulk}).
For later times the GHY term includes contributions from the future singularity and from the two asymptotic boundaries. The contributions from the cutoff surfaces at the asymptotic boundaries do not depend on time, and have the same value that they have for $t \leq t_c$. The contribution from the cutoff surface at the future singularity reads
\be
I_{\mt{surface}}^{\mt{future}}=\frac{V_{d-1}}{8\pi G_\mt{N}} \mathcal{G}(r)\left(\frac{t}{2}+r^{*}_{\infty}-r^{*}(r) \right)\Big|_{r=\epsilon_{0}}\,.
\ee
The total surface term can be written as
\be
I_\mt{surface}(t > t_c)=I_\mt{surface}(t\leq t_c)+I_{\mt{surface}}^{\mt{future}}\,.
\ee
where $I_\mt{surface}(t\leq t_c)$ is defined in equation (\ref{eq-Isurface}).
Finally, we turn to the computation of the joint terms. These terms include time-independent contributions from the two asymptotic boundaries, which are equal to the corresponding quantities for $t \leq t_c$, a vanishing contribution from the cutoff surface at the future singularity and  a contribution from the intersection of the two null boundaries of the WDW patch. The joint term can then be written as
\be
I_\mt{joint}(t>t_c)= I_\mt{joint}^\mt{bdry}+I_\mt{joint}^\mt{null}\,,
\ee
where $I_\mt{joint}^\mt{null}$ is the contribution from the intersection of the two null boundaries. This term reads
\be
I_{\mt{joint}}^\mt{null} =\frac{1}{8\pi G_\mt{N}}\int d^{d-1}x \sqrt{G}\, \bar{a}
\ee
where $\bar{a}$ is defined in terms of the left and right null vectors that parametrize the null boundaries of the WDW patch. These null vectors are given by
\be
k_{\mu}^\mt{L}=-\alpha\,\partial_{\mu}(t-r^{*})\,,\,\,\,\,\,\,k_{\mu}^\mt{R}=\alpha\,\partial_{\mu}(t+r^{*})
\label{eq-nullvectors}
\ee
In terms of $k_{\mu}^\mt{L}$ and $k_{\mu}^\mt{R}$ the quantity $\bar{a}$ can be written as
\be
\bar{a}=\log\left| \frac{1}{2}k^\mt{L}\cdot k^\mt{R}\right|=-\log\left|\frac{G_{tt}(r_{m}) }{\alpha^2}\right|\,.
\ee
Using the above expressions we can write
\be
I_{\mt{joint}}^\mt{null} =-\frac{V_{d-1}}{8\pi G_\mt{N}}\sqrt{G(r_{m})}\log\left|\frac{G_{tt}(r_{m}) }{\alpha^2}\right|\,.
\ee
where $r_m$ is given by equation (\ref{eq-rm}).
The null vectors $k_{\mu}^\mt{L}$ and $k_{\mu}^\mt{R}$ are defined in terms of an arbitrary normalization constant $\alpha$ that introduces an ambiguity in the calculation of $I_\mt{WDW}$. With the above results, the joint term can be written as
\be
I_\mt{joint}(t > t_c)=I_\mt{joint}(t \leq t_c)-\frac{V_{d-1}}{8\pi G_\mt{N}}\sqrt{G(r_{m})}\log\left|\frac{G_{tt}(r_{m}) }{\alpha^2}\right|\,.
\ee
Note that for $t > t_c$ the gravitational action calculated in the WDW patch can be written as
\be
I_\mt{WDW}(t>t_c)=I_\mt{WDW}(t \leq t_c)+\delta I\,,
\ee
where
\be
\delta I =\delta I_\mt{bulk}+\delta I_\mt{surface}+\delta I_\mt{joint}\,,
\ee
with
\begin{eqnarray}
\delta I_{\mt{bulk}} &=&I_{\mt{bulk}}(t>t_c)-I_{\mt{bulk}}(t \leq t_c) \nonumber\\
\nonumber &=&\frac{V_{d-1}}{8\pi G_\mt{N}} \int_{\epsilon_{0}}^{r_{m}} dr  \sqrt{-g} \, \mathcal{L}(r)\left[\frac{\delta t}{2}+r^{*}(r)-r^{*}(0) \right]\,,\\
\delta I_{\mt{surface}} &=&I_{\mt{surface}}(t>t_c)-I_{\mt{surface}}(t \leq t_c)\nonumber \\&=& \frac{V_{d-1}}{8\pi G_\mt{N}} \mathcal{G}(r)\frac{\delta t}{2} \Big|_{r=\epsilon_{0}}\,,\nonumber\\
\delta I_{\mt{joint}}&=&I_{\mt{joint}}(t>t_c)-I_{\mt{joint}}(t \leq t_c) \nonumber \\
&=& -\frac{V_{d-1}}{8\pi G_\mt{N}}\sqrt{G(r_{m})}\log\left|\frac{G_{tt}(r_{m}) }{\alpha^2}\right|\,.
\end{eqnarray}
It is convenient to work with the time variable $\delta t=t-t_{c}$, which is related to $r_m$ as
\begin{equation}
\frac{\delta t}{2}+r^{*}(r_{m})-r^{*}(0)=0\,.
\label{eq-deltat}
\end{equation}
Finally, the time derivative of each contribution reads
\bea
\frac{d\delta I_{\textrm{bulk}}}{dt} &=&\frac{V_{d-1}}{16\pi G_\mt{N}} \int_{\epsilon_{0}}^{r_{m}} dr  \sqrt{-g}\,\mathcal{L}(r)\,,\\
\frac{d\delta I_{\textrm{surface}}}{dt} &=& \frac{V_{d-1}}{16\pi G_\mt{N}}  \mathcal{G}\Big|_{r=\epsilon_{0}}\,,\\
\frac{d \delta I_{\textrm{joint}}}{dt} &=&\frac{V_{d-1}}{16\pi G_\mt{N}} \Big(  \sqrt{\frac{G}{G_{rr}G_{tt}}} G_{tt}' \Big)\Big|_{r=r_m}  \nonumber\\
 &+&\frac{1}{2}\sqrt{\frac{G_{tt}}{G_{rr}G}} G'\log \left|\frac{G_{tt}}{\alpha^2 } \right| \,.
\eea
The time derivative of $I_\mt{WDW}$ can then be computed as
\bea
\frac{d I_\mt{WDW}}{dt}=\frac{V_{d-1}}{16\pi G_\mt{N}} \left[  \int_{\epsilon_{0}}^{r_{m}} dr  \sqrt{-g} \,\mathcal{L}(r)+ \mathcal{G}(r)\Big|_{r=\epsilon_{0}}\right.\nonumber\\
\left. + \left( \frac{1}{2}\sqrt{\frac{G_{tt}}{G_{rr}G}}  G'\log \left|\frac{G_{tt}}{\alpha^2 } \right|+\sqrt{\frac{G}{G_{rr}G_{tt}}} G_{tt}' \right)\Big|_{r=r_m} \right]\,. \nonumber \\
\label{eq-Iwdw-t}
\eea
Therefore, the time derivative of the holographic complexity can be obtained as
\be
\frac{d\mathcal{C}_A}{dt}=\frac{1}{\pi \hbar} \frac{d I_\mt{WDW}}{dt}\,.
\ee
\subsubsection{Late time behavior} \label{sec-CAlatetime}
\begin{figure}
\centering
\setlength{\unitlength}{1cm}
\includegraphics[width=9.cm]{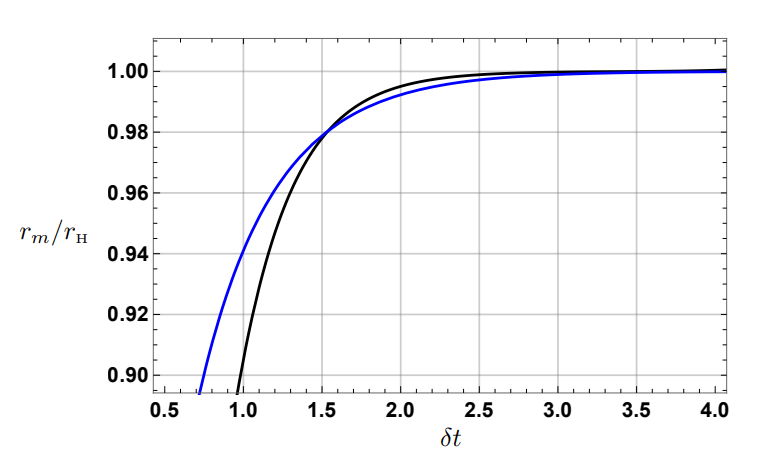} 

\caption{
$r_m / \rh$ versus $\delta t$ for the MT model (blue curve) and for the DK model (black curve). Here, for the MT model, we have fixed $\rh=1$ and $a/T=0.314$. For the DK model we fixed $B=3$ and $\rh=1$. The curves obtained for another values of these parameters are indistinguishable from the above results. }
\label{fig-rm}
\end{figure}

In this section we now apply the formula (\ref{eq-Iwdw-t}) for the MT and DK models to study the late time behavior of the time-derivative of $\mathcal{C}_A$. We first observe that, at later times, $r_m$ approaches $\rh$. This can be seen in figure \ref{fig-rm}, where we plot $r_m$ versus $\delta t$.

Therefore, the late time behavior of $dI_\mt{WDW}/dt$ is obtained by taking the limit $r_m \rightarrow \rh$ in the equation (\ref{eq-Iwdw-t})
\begin{eqnarray}
\frac{d I_\mt{WDW}}{dt}&=&\frac{V_{d-1}}{16\pi G_\mt{N}}\Bigg[ \int_{\epsilon_{0}}^{\rh} dr  \sqrt{-g}\,\mathcal{L}(r)+\sqrt{\frac{G}{G_{rr}G_{tt}}} G_{tt}' \Big|_{r=\rh} \nonumber\\
  &+&\sqrt{\frac{G_{tt}G}{G_{rr}}}\left(\frac{G_{tt}'}{G_{tt}}+\frac{G'}{G} \right)\Big|_{r=\epsilon_{0}} \Bigg]\,.
\label{eq-Iwdw-tlarge}
\end{eqnarray}
We have checked that the same late-time result for $\frac{d\mathcal{C}_A}{dt}$ can be obtained by following the approach developed by Brown et al \cite{CA-2}. See appendix \ref{appB}.
\subsubsection*{Results for the MT model}
Substituting the metric functions $G_{mn}(r)$ and the on-shell Lagrangian $\mathcal{L}(r)$ for the MT model and expanding the above contributions for small anisotropies, we obtain
\bea
 \int_{\epsilon_{0}}^{\rh} dr  \sqrt{-g}\,\mathcal{L}(r)=-2\rh^4-\frac{5}{6} \rh^2 a^2 \log2 +\mathcal{O}(\epsilon_0^4 \log\epsilon_0)\,,\nonumber \\
\sqrt{\frac{G_{tt}G}{G_{rr}}}\left(\frac{G_{tt}'}{G_{tt}}+\frac{G'}{G} \right)\Big|_{r=\epsilon_{0}} =4\rh^4+\frac{1}{6} \rh^2 a^2 \left(5\log2-1\right)\nonumber\\+\mathcal{O}(\epsilon_0^4 \log\epsilon_0)\,,\nonumber \\
\sqrt{\frac{G}{G_{rr}G_{tt}}} \partial_r G_{tt} \Big|_{r=\rh}=4\rh^4+\frac{1}{3} \rh^2 a^2 \left(10\log2-1\right)\,.\nonumber\\
\eea
By summing the above contributions and taking the limit $\epsilon_0 \rightarrow 0$, we find
\be
\frac{d I_\mt{WDW}}{dt}= \frac{V_3}{16\pi G_\mt{N}}\left( 6 \rh^4+\frac{\rh^2 a^2}{2}   \left(5\log(2)-1 \right) \right)= 2 M(a)\,,
\label{eq-CAlatetime2}
\ee
where the mass of the black brane $M(a)$ is given by equation (\ref{eq-mass}). Therefore, the late time behavior of the time derivative of holographic complexity reads
\be
\frac{d\mathcal{C}_A}{dt}=\frac{2M(a)}{\pi \hbar}\,,
\label{eq-CAlatetime}
\ee
which saturates the Lloyd's bound. 

\subsubsection*{Results for the DK model}

For the DK model, we obtain the following results
\be
\frac{dI_\mt{WDW}}{dt}=\frac{V_3}{16 \pi G_\mt{N}}  \left( \delta \dot{I}_\mt{bulk}+\delta \dot{I}_\mt{surface}+\delta \dot{I}_\mt{joint} \right)
\ee
where the contributions from bulk, surface and joint terms are given by
\bea
\delta \dot{I}_\mt{bulk}&=& \int_{0}^{\rh}dr \sqrt{-g}\,\mathcal{L}(r)=-6 B \, \rh^2\,, \nonumber\\
\delta \dot{I}_\mt{surface}&=&\sqrt{\frac{G_{tt}G}{G_{rr}}}\left( \frac{G_{tt}'}{G_{tt}}+\frac{ G'}{G} \right)\Big|_{r=0}=6 B \, \rh^2\,,\nonumber\\
\delta \dot{I}_\mt{joint}&=&\sqrt{\frac{G}{G_{tt}G_{rr}}} G_{tt}'\Big|_{r=\rh}=6 B \, \rh^2\,.
\eea
With the above results, the late-time rate of change of holographic complexity reads
\be
\frac{d\mathcal{C}_A}{dt}=\frac{1}{\pi}\frac{dI_\mt{WDW}}{dt}=\frac{V_3}{16 \pi G_\mt{N}} \times 6 B \, \rh^2=2 M_B\,,
\ee
which precisely saturates the Lloyd's bound. This provides another example where, despite the anisotropy, the Lloyd's bound is still respected.

\subsubsection{Full time behavior}
In this section we study the full time behavior of holographic complexity for the MT and DK models. We numerically solve the equation (\ref{eq-deltat}) to find $r_m$ as a function of $\delta t$ and then we use the result in equation (\ref{eq-Iwdw-t}) to obtain $I_\mt{WDW}$ as a function of $\delta t$. 

\subsubsection*{Results for the MT model}

\begin{figure}

\includegraphics[width=8.5cm]{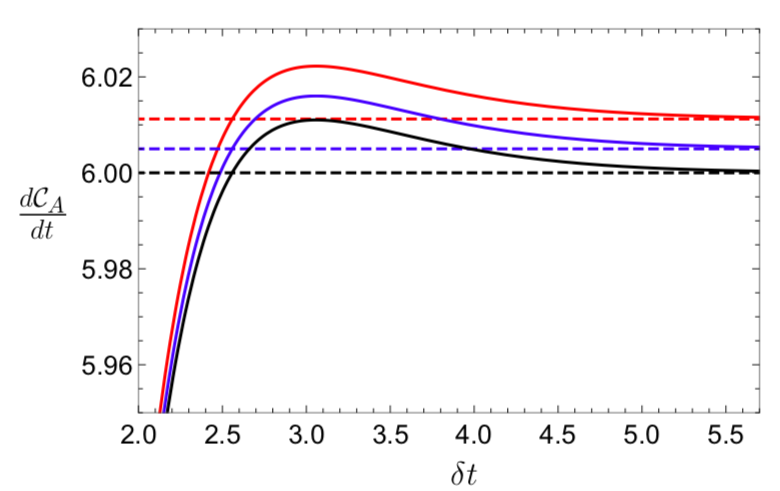} 

\caption{ \small
The time dependence of holographic complexity calculated with the CA proposal. The curves correspond to: $(a,\rh,2M)=(0,1,6)$ (black curves), $(a,\rh,2M)=(0.1,0.9997,6.005)$ (blue curves) and $(a,\rh,2M)=(0.15,0.9993,6.011)$ (red curves). The continuous curves represent the results (in units of $V_3/(16\pi^2\, \hbar \,G_\mt{N})$) for the time derivative of holographic complexity, while the dashed horizontal lines represent $2M$. We fix the normalization of the null-vector in equation (\ref{eq-nullvectors}) by taking $\alpha=0.1$. The qualitative behavior is the same for other values of $\alpha$.}
\label{fig-Iversust}
\end{figure}

The geometry in the MT model is controlled by the dimensionless parameter $a\rh$, where $a$ is the parameter of anisotropy.
The values of $(a,\rh,M)$ for which we study the complexity growth are shown in table 1, and they were chosen such that the temperature is fixed as we increase the anisotropy. In this table we can see that the black brane's mass increases as we increase $a$ while keeping $T$ fixed. Figure \ref{fig-Iversust} shows the time dependence of the gravitational action in the WDW patch for the choice of parameters presented in table 1. The behavior of $d\mathcal{C}_A/dt$ is qualitatively similar to the behavior observed in isotropic systems. The anisotropy increase the mass of the black brane and its effects on the rate of change of complexity seem to be just a vertical shift in the curves of $d\mathcal{C}_A/dt$ versus $t$.

\begin{table}
\begin{tabular}{ |c|c|c|c| } 
\hline
anisotropy parameter & $\rh$ such that  $T=1/\pi$ & $2 M$ \\
\hline
$0.00$ & $1.0000$ & $6.000$ \\ 
$0.10$ &  $0.9997$ & $6.005$ \\ 
$0.15$ &  $0.9993$ & $6.011$\\ \hline
\end{tabular}
\caption{black brane's mass, measured in units of $V_3/(16\pi G_\mt{N})$, for several values of $a$ and $\rh$.
Here we chosen $\rh$ such that the Hawking temperature is fixed $T=1/\pi$.}
\end{table}

\subsubsection*{Results for the DK model}
\begin{figure}
\includegraphics[width=8.5cm]{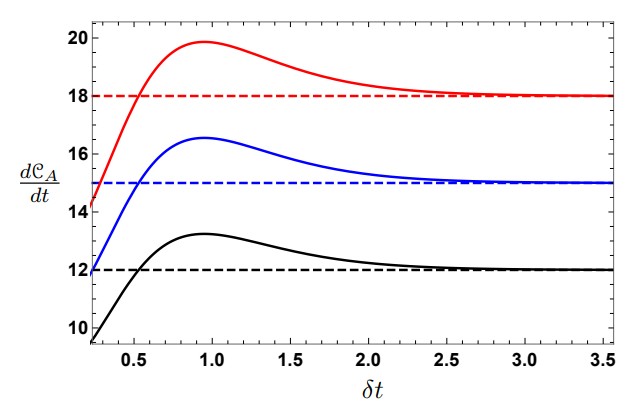} 

\caption{ \small Full time behavior of holographic complexity of magnetic black branes for different values of the magnetic field. The curves correspond to $B=2$ (black curves), $B=2.5$ (blue curves) and $B=3$ (red curves). The continuous curves corresponds represent $\dot{\mathcal{C}_A}$, while the dashed horizontal lines represent $2M_B$. The results are given in units of $V_3/(16\pi^2 G_\mt{N})$. We fixed the normalization of the null vector in equation (\ref{eq-nullvectors}) by taking $\alpha=1.3$. The qualitative behavior is the same for other values of $\alpha$.}
\label{fig-IBversust}
\end{figure}

The geometry of the DK model is controlled by the dimensionless parameter $B/T^2$, where $B$ is the intensity of the magnetic field, while $T$ is the black brane's temperature \footnote{Differently from the MT model, the temperature of the DK solution does not depend on the intensity of the magnetic field (see (\ref{T-DK})), so we don't need to change $\rh$ to keep the temperature fixed while we vary $B$.}.
In figure \ref{fig-IBversust} we show the full time behavior of the rate of change of complexity for different intensities of the magnetic field. Just like in the MT model, there is a violation of Lloyd's bound at early times, and the result approach the bound from above at later times. Moreover, the net effect of the magnetic field is just a vertical shift in the curves of $\dot{\mathcal{C}_A}$  versus $\delta t$.

\section{The charged case}\label{sec-charged}
In this section we use the CA proposal to study the rate of change of complexity of charged anisotropic black branes. In particular, we consider the type IIB supergravity solution found in \cite{Cheng:2014qia,Cheng:2014sxa}. This solution is basically an extension of the MT solution to the charged case, therefore, for conceptual clarity, we will refer to it as the {\it charged MT model}. The action of this model reads
\bea
S&=&\frac{1}{16\pi G_\mt{N}}  \int_{\mathcal{M}}  d^{5}x\sqrt{-g}\Big[R+\frac{12}{L^2}-\frac{1}{2}\left( \partial\phi\right)^2 \\
\nonumber &-&\frac{1}{2}e^{2\phi}\left( \partial\chi \right)^2-\frac{1}{4}F_{mn}F^{mn} \Big]+S_\textrm{GH}\,.
\label{eq-actionCharged}
\eea
The solution in Einstein frame takes the form given by (\ref{eq-metric}) and (\ref{eq-axionDilaton}), with the metric functions given by
\bea
\mathcal{F}&=&1-\frac{\rh^4}{r^4}+\left[\left( \frac{\rh}{r}\right)^6- \left( \frac{\rh}{r}\right)^4\right]q^2 + a^2 \mathcal{F}_2(r,q)+\mathcal{O}(a^4) \nonumber\\
B&=&1+a^2 \mathcal{B}_2(r,q)+\mathcal{O}(a^4) \nonumber\\
\mathcal{H}&=&e^{-\phi(r)}\,,\,\, \text{with}\,\,\phi = a^2 \phi_2(r,q)+\mathcal{O}(a^4) 
\eea
where the functions $\mathcal{F}_2(r,q), \mathcal{B}_2(r,q)$ and $\phi_2(r,q) $ now depend on the charge parameter $q$, which is related to the black brane's charge as $q \equiv \frac{Q}{\rh^3 2 \sqrt{3}}$.  Here $Q$ is the black brane's charge in units of $V_3/(16\pi G_\mt{N})$. For small values of $q$, one can find an analytic solution of the form \cite{Cheng:2014qia} 
\bea
\mathcal{F}_2(r,q)=f_0(r)+f_2(r)q^2+\mathcal{O}(q^4) \,,\nonumber\\
 \mathcal{B}_2(r,q)=b_0(r)+b_2(r)q^2+\mathcal{O}(q^4)\,,\nonumber\\
\phi_2(r,q)=\varphi_0(r)+\varphi_2(r)q^2+\mathcal{O}(q^4)\,,
\eea
where the $\mathcal{O}(q^0)$ terms are
\bea
f_0(r)&=&-\frac{1}{24 r^4 \rh^2} \Bigg[8r^2\rh^2 - 2\rh^2(4+5\log 2)\nonumber \\
&+&(3r^4+7\rh^4)\log \left(1+\frac{\rh^2}{r^2} \right) \Bigg] \,, \nonumber \\
b_0(r)&=&-\frac{1}{24 \rh^2} \left[\frac{10 \rh^2}{r^2+\rh^2}+\log \left(1+\frac{\rh^2}{r^2} \right) \right] \,, \nonumber\\
\varphi_0(r)&=&-\frac{1}{4\rh^2} \log \left(1+\frac{\rh^2}{r^2} \right)\,,
\eea
while second order terms are given by
\bea
f_2(r)&=&\frac{1}{24 r^6\rh^2(r^2+\rh^2)}\Bigg[ 6r^6\rh^2+\rh^8\nonumber \\ &+&r^4\rh^4 (25 \log2-12)
r^2\rh^6(25\log2-1)\nonumber \\&-&(r^2+\rh^2)(6r^6+7r^2\rh^4+12\rh^6)\log\left( 1+\frac{\rh^2}{r^2}\right) \Bigg]\,, \nonumber\\
b_2(r)&=&-\frac{2r^4+3r^2\rh^2+11\rh^4}{24 r^2(r^2+\rh^2)^2} +\frac{1}{12\rh^2} \log\left( 1+\frac{\rh^2}{r^2} \right)\,, \nonumber\\
\varphi_2(r)&=&-\frac{1}{4r^2}-\frac{1}{4(r^+\rh^2)}+\frac{1}{2}\log\left( 1+\frac{\rh^2}{r^2} \right)\,.
\eea

The field-strength and the associated chemical potential are given by
\bea
F&=&-Q\sqrt{\mathcal{B}}e^{3\phi/4}\frac{1}{r^3} dt \wedge dr\,, \nonumber\\
\mu &=& \frac{Q}{2} \left(1-\frac{5a^2}{24\rh^2}\log2 \right)\,.\nonumber\\
\label{eq-Fmu}
\eea
The charged black brane's mass is given by
\bea
M(a,q)&=& 3 \rh^4+\frac{a^2\rh^2}{8} \Big(-2+10\log2 \Big)\nonumber \\
&+&q^2\Big[3 \rh^4 -\frac{5a^2\rh^2}{8} (-3+5\log2)\Big] \,.
\eea

\subsection{Rate of change of complexity}
\begin{figure}
\centering
 \includegraphics[width=8.5cm]{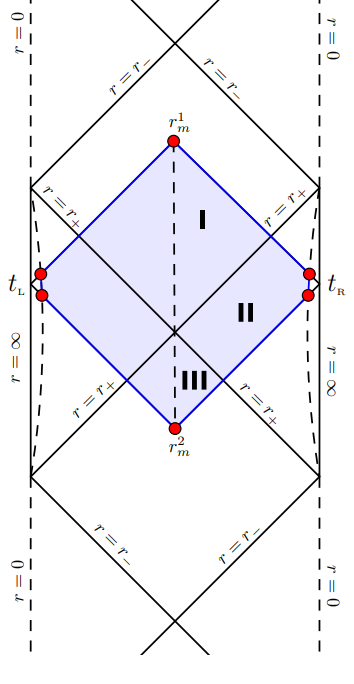}      
     
\caption{ Penrose diagram of a charged asymptotically AdS black hole. We split half of the WDW patch into three regions, I, II and III. At later times, the future corner, $r_m^1$, approaches the inner horizon, $r_{-}$, while the past corner, $r_m^2$, approaches the outer horizon, $r_{+}$. }
\label{fig-PenroseCharged}
\end{figure}

The rate of change of complexity can be calculated as before, by considering the on-shell action evaluated on the WDW patch, with the difference that now the Penrose diagram is modified by the fact that the black hole is charged. See figure \ref{fig-PenroseCharged}. The total action is given by a sum of four terms: the bulk contribution, the surface contribution,  the joint contribution, and a boundary term for the Maxwell field. 

Let us first evaluate the sum of the bulk and Maxwell contributions. Using the equations of motion that result from the action (\ref{eq-actionCharged}), it is easy to show that the on-shell lagrangian density is given by
\be
\mathcal{L}(r)= -8-\frac{1}{6}F_{mn}F^{mn}\,.
\ee
However, this on-shell Lagrangian density can be affected by the presence of a non-zero boundary term for the Maxwell field. For a gauge field action of the form
\be
I_\mt{Maxwell}=-\frac{1}{4g^2}\int_{\mathcal{M}} d^{d+1}x \sqrt{-g}F_{mn} F^{mn}\,,
\ee
where $g$ is the gauge coupling parameter \footnote{In the charged MT model, one has $g^2 = 16 \pi G_\mt{N}$.}. The corresponding boundary term can be written as \cite{Goto:2018iay}
\be
I_\mt{Maxwell}^\mt{bdry}=\frac{\gamma}{g^2}\int_{\partial \mathcal{M}} d\Sigma_m F^{mn}A_m\,,
\label{eq-MaxB}
\ee
where $\gamma$ is an arbitrary parameter that affects the late time behavior of complexity. Later, we are going to fix this parameter by requiring consistency with the uncharged case.  Using the equations of motion, one can show that
\be
I_\mt{Maxwell}^\mt{bdry}\Big|_\mt{on-shell}=\frac{\gamma}{2g^2}\int_{\mathcal{M}} d^{d+1}x \sqrt{-g}F_{mn} F^{mn}\,.
\ee
Therefore, taking into account the above contribution, the on-shell Lagrangian density becomes
\be
\mathcal{L}(r)= -8-\frac{1-2\gamma}{6}F_{mn}F^{mn}\,.
\ee
We now proceed to the evaluation of the bulk action corresponding to the above on-shell Lagrangian density. The WDW patch is shown in Figure \ref{fig-PenroseCharged}. The future and past corners are denoted as $r_m^1$ and $r_m^2$, respectively. The $r-$coordinate of these points satisfy the following relations
\be
\frac{t}{2}+r_{\infty}^*-r^*(r_m^1)=0\,,\,\,\,\,\frac{t}{2}-r_{\infty}^*-r^*(r_m^2)=0
\label{eq-rm12definition}
\ee
The time derivative of the above relations implies
\be
\frac{dr_m^{1,2}}{dt}=\pm \frac{\sgn{G_{tt}}}{2}\sqrt{\frac{G_{tt}}{G_{rr}}}\Big|_{r=r_m^{1,2}}\,.
\ee
As before, we calculate the bulk contribution for half of the WDW patch, and then we multiply the final result by two. The contributions for regions I, II and III are given by
\bea
I_\mt{bulk}^\mt{I}&=&\frac{V_{d-1}}{16\pi G_\mt{N}} \int_{r_m^1}^{r_{+}} dr \sqrt{-g} \mathcal{L}(r) \left( \frac{t}{2}+r_{\infty}^*-r^*(r)\right)\,, \nonumber\\
I_\mt{bulk}^\mt{II}&=&\frac{V_{d-1}}{8\pi G_\mt{N}} \int_{r_+}^{r_\mt{max}} dr \sqrt{-g} \mathcal{L}(r) \Big( r_{\infty}^*-r^*(r)\Big)\,,\nonumber\\
I_\mt{bulk}^\mt{III}&=&\frac{V_{d-1}}{16\pi G_\mt{N}} \int_{r_m^2}^{r_{+}} dr \sqrt{-g} \mathcal{L}(r) \left(- \frac{t}{2}+r_{\infty}^*-r^*(r)\right)\,. \nonumber \\
\eea
By taking the time derivative of the above expressions and using (\ref{eq-rm12definition}) one can check that
\be
\frac{dI_\mt{bulk}}{dt}=-\frac{V_{d-1}}{16\pi G_\mt{N}} \int_{r_m^1}^{r_m^2}dr \sqrt{-g} \Big(8+\frac{1-2\gamma}{6} F^2\Big)\,.
\ee
We now turn to the evaluation of the joint terms, corresponding to the future and past corners (the red dots in Figure \ref{fig-PenroseCharged}). These contributions are given by
\be
I_\mt{joint}^{1,2}=-\frac{V_{d-1}}{8\pi G_\mt{N}} \sqrt{G(r_m^{1,2})} \log \Big| \frac{G_{tt}(r_m^{1,2})}{\alpha^2} \Big|
\ee
whose time derivative gives
\bea
\frac{d I_\mt{joint}^{1,2}}{dt}=-\frac{V_{d-1}}{16\pi G_\mt{N}} \Bigg( \sqrt{\frac{G_{tt}}{G_{rr}G}}G'(r) \log \Big| \frac{G_{tt}}{\alpha^2} \Big| \nonumber \\
+\, \sqrt{\frac{G}{G_{rr}G_{tt}}} G_{tt}'(r) \Bigg) \Bigg|_{r=r_m^{1,2}}\,.
\eea
Finally, the surface contributions coming from the two asymptotic boundaries do not depend on time, and hence do not contribute the rate of change of complexity. Adding the above contributions, we find
\bea
\pi \frac{d \mathcal{C}_A}{dt}=\frac{V_{d-1}}{16\pi G_\mt{N}} \Bigg[ \int_{r_m^1}^{r_m^2} dr \sqrt{-g} \left( 8+\frac{1-2\gamma}{6}F^2 \right)\nonumber \\
-\Bigg( \sqrt{\frac{G_{tt}}{G_{rr}G}}G' \log \Big| \frac{G_{tt}}{\alpha^2}\Big|+ \sqrt{\frac{G}{G_{rr}G_{tt}}} G_{tt}'  \Bigg)\Big|_{r_m^2}^{r_m^1} \Bigg]
\eea
At late times, $r_m^1 \rightarrow r_{-}$ and $r_m^2 \rightarrow r_{+}$, we find
\bea
\pi  \frac{d \mathcal{C}_A}{dt}=\frac{V_{d-1}}{16\pi G_\mt{N}} \Bigg[ \int_{r_+}^{r_-} dr \sqrt{-g} \left( 8+\frac{1-2\gamma}{6}F^2\right)\nonumber \\
-\, \sqrt{\frac{G}{G_{rr}G_{tt}}} G_{tt}'(r) \Big|_{r_+}^{r_-} \Bigg]\,.
\label{eq-CAqlatetime}
\eea
In order to evaluate the above formula for the charged MT model, we parametrize the outer and inner horizon as
\be
r_{+}=\rh\,,\,\,\,\,r_{-}=\rh q \left[1+\frac{a^2}{48 \rh^2} \left( 5 \log \frac{q^2}{4}-1\right) \right]\,.
\label{eq-horizons}
\ee
Specializing (\ref{eq-CAqlatetime}) for the charged MT model, and using (\ref{eq-horizons}), we find
\bea
\pi  \frac{d \mathcal{C}_A}{dt}&=&\frac{V_{3}}{16\pi G_\mt{N}} \rh^4 (3-2\gamma) \Bigg[ 2(1-q^2)+\nonumber \\
&+&\frac{a^2}{12\rh^2} \Big(1+5q^2(-1+\log2)+10\log2 \Big) \Bigg].
\label{eq-resultgamma}
\eea
The result depends on the arbitrary parameter, $\gamma$. This parameter, however, can be fixed by requiring the $q \rightarrow 0$ limit of (\ref{eq-resultgamma}) to be consistent with the uncharged case. This can be done by setting $q = 0$ in (\ref{eq-resultgamma}) and choosing $\gamma$ such that the final result matches (\ref{eq-CAlatetime2}) \footnote{Such tunning of $\gamma$ was also observed to be necessary in~\cite{Jiang:2019qea}.}. By doing that, we find
\be
\gamma= \frac{3a^2}{16\rh^2}\,.
\ee
Notice that $\gamma=0$ for $a=0$, which means that the contribution of the Maxwell boundary term is zero in the isotropic case. 
With this choice for $\gamma$, the final result reads
\bea
\pi  \frac{d \mathcal{C}_A}{dt}&=&\frac{V_{3}}{16\pi G_\mt{N}} \Bigg( 6(1-q^2)\rh^4 \nonumber \\
&+&\frac{a^2\rh^2}{2}\Bigg[-1+5\log2+q^2\left(-1+\frac{5}{2}\log2\right) \Bigg] \Bigg) \,.\nonumber\\
\eea
By construction, the $q \rightarrow 0$ limit of the above result is consistent with the result for the uncharged case, given in (\ref{eq-CAlatetime2}). Furthermore, the $a \rightarrow 0$ limit is consistent with previous results reported in the literature. See, for instance, (4.27) of \cite{Carmi-2017}.

Now let us discuss our result in the light of the bound proposed in \cite{CA-2}, according to which the natural bound for states at a finite chemical potential is
\be
 \pi  \frac{d \mathcal{C}_A}{dt} \leq 2 \left(M-\mu \mathcal{Q} \right)- 2 \left(M-\mu \mathcal{Q} \right)_\mt{gs}\,,
\label{eq-boundQ}
\ee
where the second term correspond to the ground state (gs) value of $\left(M-\mu \mathcal{Q} \right)$. As the charged MT model does not have an extremal limit, the ground state is  the vacuum solution ($M=\mathcal{Q}=0$).
Using the formula (\ref{eq-Fmu}) for the chemical potential, and taking into account that the black brane's charge is $\mathcal{Q} =\frac{V_{3} Q}{16\pi G_\mt{N}}$, we can see that
\be
\pi  \frac{d \mathcal{C}_A}{dt} - 2 \left(M-\mu \mathcal{Q} \right) =\frac{a^2q^2\rh^2}{4}\left(-17+20\log2 \right)<0\,
\ee
which shows that the bound (\ref{eq-boundQ})  can be saturated in the $a \rightarrow 0$ limit, but it is no longer saturated once we turn on the anisotropy parameter.

\section{Discussion} \label{sec-disc}

We have used the CA conjecture to study the time-dependence of holographic complexity for three anisotropic black brane solutions, namely, the MT model, the DK model, and the charged MT model.

\subsection*{MT model}

The MT solution is dual to the $\mathcal{N}=4$ SYM theory deformed by a position-dependent theta-term that breaks isotropy and conformal invariance. The background geometry is controlled by the ratio $a/T$, where $a$ is the parameter of anisotropy, and $T$ is the Hawking temperature. 

Similarly to the case of isotropic systems, the rate of change of complexity in anisotropic systems is zero for $t \leq t_c$, and it is non-zero for $t>t_c$, with this critical time given by equation (\ref{eq-tc}). Figure \ref{fig-tc} shows the behavior of $t_c$ as a function of the anisotropy parameter. In this figure we consider increasing values of the anisotropy parameter, while keeping fixed the temperature. As compared with an isotropic system with the same temperature, the holographic complexity of anisotropic systems remains constant for a shorter period, i.e., the effect of the anisotropy is to reduce $t_c$.

In section \ref{sec-CAlatetime} we study the late-time behavior of the holographic complexity and find an expression for $dI_\mt{WDW}/dt$ in terms of the metric functions. See equation (\ref{eq-Iwdw-tlarge}). For simplicity, let us first consider the isotropic case, in which $a=0$. In this case the MT solution reduces to the five-dimensional black brane solution that is dual to the undeformed $\mathcal{N}=4$ SYM theory. From previous works \cite{CA-2,Carmi-2017,Kim-2017a}, we know that the Lloyd's bound should be respected in this case. As we turn on a small anisotropy parameter, all the metric functions get corrections up to the second order in $a$ and this leads to a larger black brane's mass (see equation (\ref{eq-mass})). In this case, we expect the formula (\ref{eq-Iwdw-tlarge}) to provide the result for $a=0$, plus corrections up to the second order in the anisotropy parameter. Applying our formulas for the MT model we find that the late time rate of change of complexity matches the Lloyd's bound, i.e., $d\mathcal{C}_A/dt=2M(a)/\pi \hbar$. This is a highly non-trivial match, because it means that the anisotropy increases the value of $2M$ and the late time value of $dI_\mt{WDW}/dt$ precisely in the same amount.

The full-time behavior of $d\mathcal{C}_A/dt$ can be seen in figure  \ref{fig-Iversust}. The results share a lot of similarities with the previous results obtained for isotropic systems \cite{Carmi-2017}. In particular, $d\mathcal{C}_A/dt$ violates the Lloyd's bound at initial times, and approaches this bound (from above) at later times. In this figure we consider increasing values of the anisotropy parameter, while keeping fixed the temperature. The resulting black brane's mass increases as we increase the anisotropy parameter, and the overall effect of the anisotropy is a vertical upward shift \footnote{We have checked that the curves of $\frac{1}{M(a)}\frac{d\mathcal{C}_A}{dt}$ versus $\delta t$ are indistinguishable for different (and small) anisotropies. This confirms that the basic effect of the anisotropy is a upward shift in the curves of $\frac{d\mathcal{C}_A}{dt}$ versus $\delta t$. We thank Alberto G\"uijosa for suggesting this comparison.} in the curves of  $d\mathcal{C}_A/dt$ versus $\delta t$. At later times, the difference between the anisotropic and isotropic results is proportional to the difference in pressures in the transverse and longitudinal directions, namely
\be
\frac{d\mathcal{C}_A}{dt}=\frac{2M(0)}{\pi \hbar}+\frac{V_3}{\pi \hbar}(P_{xy}-P_z)+\mathcal{O}(a^4)\,.
\ee
This can be seen from equations (\ref{eq-CAlatetime}) and (\ref{eq-compMass}).

\subsection*{DK model}

The behavior of holographic complexity in the MT model is very similar to the behavior observed in magnetic branes. By using the CA conjecture, we studied the time behavior of holographic for the magnetic black brane solution found by D'Hoker and Kraus in \cite{DHoker:2009mmn}. In this model one introduces a constant magnetic field that breaks the rotational symmetry of the background. The geometry is controlled by the ratio, $B/T^2$, between the magnetic field and the temperature squared. For very large values of values of $B/T^2$, this system has a simple solution, which is given in (\ref{eq-solB}). For this configuration, the Lloyd's bound is violated at early times, but it is saturated at later times. This provides another example of a system that breaks the rotational symmetry without violating the Lloyd's bound at later times. This should be contrasted with the bound for $\eta/s$, which is known to be violated in anisotropic systems. This suggests that the violation of Lloyd's bound \cite{Couch-2017, Alishahiha-2018,Swingle-2017,An-2018} in the case of neutral black holes is not due to anisotropy, but rather to the presence of a conformal anomaly. As neither the D'Hoker \& Kraus nor the Mateos \& Trancanelli model display a conformal anomaly (up to second order in the anisotropy), this would explain why the Lloyd's bound is not violated in these two models. We are currently investigating whether this last statement is true.

\subsubsection*{Charged anisotropic black branes}
In section \ref{sec-charged} we use the CA conjecture to study the late-time behavior of holographic complexity for a generalization of the MT model to the charged case. In this case the geometry is not only controlled by the parameter $a/T$, but also by the dimensionless charge parameter $q$. Following \cite{Goto:2018iay} we consider the inclusion of a Maxwell boundary term (see equation (\ref{eq-MaxB})), which introduces an arbitrary parameter $\gamma$ that affects the late-time rate of change of complexity. This new boundary term turned out to be necessary to make the $q \rightarrow 0$ limit of the final result consistent with the result for neutral anisotropic black branes. We find $\gamma=\frac{3a^2}{16\rh^2}$, which suggests that the Maxwell boundary term is generically necessary when we have non-trivial matter fields besides the Maxwell field. Having fixed the value of $\gamma$, we find that the charged MT model respects the bound (\ref{eq-boundQ}) proposed in \cite{CA-2}.

\subsubsection*{Conclusions and Future directions}

We have considered three different models in which matter fields break the rotational symmetry, and we studied how this affects the holographic complexity. In neutral black holes, the formula for holographic complexity only depends on the metric components, having no explicit dependence on the matter fields (the on-shell Lagrangian is just a constant). In other words, the matter fields only affect the holographic complexity through their effect on the geometry. This should be contrasted with the charged case, in which the electric charge appears explicitly in the formulas for the holographic complexity, as well as in the metric components.

We have observed that the holographic complexity of anisotropic systems increases as compared to isotropic systems at the same temperature. This happens because the matter fields increase the mass of the black hole and, for the Lloyd's bound to be respected, the holographic complexity also has to increase.

We have studied the effects of anisotropy on the complexity growth considering the case of small anistropies. Our results are valid up to $\mathcal{O}(a^2)$. It would be interesting to extend our results to higher anisotropies, because in this case the MT model displays a conformal anomaly \footnote{In the MT model the conformal anomaly appears at order $\mathcal{O}(a^4)$.}, which might cause a violation of the Lloyd's bound. Besides that, the MT gravitational solution can be thought of as describing a renormalization group (RG) flow from a AdS geometry in the ultraviolet (UV) to a Lifshitz geometry in the infrared (IR). The parameter controlling this transition is the ratio $a/T$, which is small close to the UV fixed point and large close to the IR fixed point. It would be interesting to study how the complexity growth behaves under this RG flow. Moreover, as Lifshitz geometries were known to violate the Lloyd's bound \cite{An-2018}, we expect such a violation to occur in the MT model at higher anisotropies.

Another interesting extension of this work would be to study the effects of the anisotropy in the holographic complexity calculated using the CV conjecture. Although this calculation is relatively easy for isotropic systems \cite{CV-2,Carmi-2017,Kim-2017a}, the extension for anisotropic systems is non-trivial, because in this case the ansatz for the maximum volume surface is more complicated, preventing the use of the techniques used in \cite{CV-2,Carmi-2017,Kim-2017a}. More specifically, the volume functional of the co-dimension one surface can be generically written as
\be
\mathcal{V} = \int d^d \sigma \sqrt{\det (g_{ab})}\,,
\label{eq-volume}
\ee
where $\sigma_a$ and $g_{ab}= \partial_a X^m \partial_b X^n G_{mn}$ are the coordinates and the induced metric along the surface, respectively. Here $X^m(\sigma_a)$ are the embedding functions describing the surface, while $G_{mn}$ are the metric components of the background geometry. In isotropic geometries, one can assume the ansatz $X^m=(v(\lambda),r(\lambda),x,y,x)$, where $v=t+r_*$. The $xyz$-rotational symmetry of this ansatz results in a simple form for the volume functional (\ref{eq-volume}), which can be easily extremized. In anisotropic systems, one no longer has this rotational symmetry, because $G_{xx} \neq G_{zz}$, and that results in a more complicated form for $X^m$ and $\mathcal{V}$.

\noindent {\bf Acknowledgements:} We are grateful to Hugo Marrochio, Hesam Soltanpanahi and Alberto G\"uijosa for very useful discussions and comments. We also thank Hamid Rajaian for useful correspondence and for comments on the manuscript. We are indebted to an anonymous referee for helpful suggestions and comments. MMQ is supported by Institute for Research in Fundamental Sciences(IPM). VJ and YDO were supported by Mexico's National Council of Science and Technology (CONACyT) grant CB-2014/238734. VJ is also partially supported  by the Basic Science Research Program through the National Research Foundation of Korea(NRF) funded by the Ministry of Science, ICT \& Future Planning(NRF2017R1A2B4004810) and GIST Research Institute(GRI) grant funded by the GIST in 2019.
\appendix

\section{Joint terms at the $r=r_\mt{max}$ and $r=\epsilon_0$ cutoff surfaces} \label{appA}
In this appendix we briefly review how to calculate the joint terms at the asymptotic boundaries and at the singularities. We show that the contributions from the asymptotic boundaries are time-independent, while the contributions from the singularities vanish.

A joint term for a corner involving the connection of at least one null surface has the form \cite{Carmi-2016}
\be
I_\mt{joint}=\frac{1}{8\pi G_\mt{N}} \int d^{d-1}x\, \sqrt{\sigma}\, \bar{a}
\ee
where $\sigma$ is the induced metric on the surfaces and $\bar{a}$ is defined as
\[
 \bar{a} = \pm
  \begin{cases} 
   \log|k \cdot n^{(t)}| & \text{for spacelike-null joints }\\
   \log|k \cdot n^{(s)}| & \text{for timelike-null joints } \\
   \log|k^{+} \cdot k^{-}/2| & \text{for null-null joints}
  \end{cases}
\]
where $k^+$ and $k^-$ are outward directed null normal vectors, while $n^{(t)}$($n^{(s)}$) are outward directed timelike (spacelike) normal vectors. The overall sign depends on the orientation of the normal vectors. For more details, see the appendix A of \cite{Carmi-2016}. 
The relevant normal vectors can be written as
\bea
n^{(t)}_{\mu}&=&(n^{(t)}_t,n^{(t)}_r,n^{(t)}_i)=(0,\sqrt{-G_{rr}(\epsilon_0)},0)\,,\\
n^{(s)}_{\mu}&=&(n^{(s)}_t,n^{(s)}_r,n^{(s)}_i)=(0,\sqrt{G_{rr}(r_\mt{max})},0)\,,\\
k^{\pm}_{\mu}&=&\pm \alpha\, \partial_{\mu}(t \pm r^*)\,.
\eea

With the above definitions, the joints term coming from the singularities can be written as
\begin{eqnarray}
I_\mt{joint}^\mt{sing}&=&\frac{1}{8\pi G_\mt{N}} \int d^{d-1}x \,\sqrt{\sigma}\,\log|k \cdot n^{(t)}| \\
\nonumber &=& -\frac{V_{d-1}}{8\pi G_\mt{N}}G(r) \log|G_{tt}(r)| \Big|_{r=\epsilon_0}\,.
\end{eqnarray}
For the MT model, one can show that $I_\mt{joint}^\mt{sing} \sim \epsilon_0^3 \log \epsilon_0$. Therefore, the contribution from this joint term vanishes in the limit $\epsilon_0 \rightarrow 0$.
The joint terms coming from the asymptotic boundaries are given by
\begin{eqnarray}
I_\mt{joint}^\mt{bdry}&=&\frac{1}{8\pi G_\mt{N}} \int d^{d-1}x \,\sqrt{\sigma}\,\log|k \cdot n^{(s)}| 
\\ \nonumber  &=& \frac{V_{d-1}}{8\pi G_\mt{N}}G(r) \log|G_{tt}(r)| \Big|_{r=r_\mt{max}}\,.
\end{eqnarray}
For the MT model $I_\mt{joint}^\mt{bdry}$ gives rise to a divergent contribution that is independent of time, because it only depends on quantities calculated on the outside of the black brane, and this region has a time-translation symmetry. Therefore, this term do not contribute to the rate of change of holographic complexity.

\section{Comparison with Brown et al} \label{appB}
The CA conjecture was proposed by Brown et al in \cite{CA-1,CA-2}. In those papers the authors find a clever way of calculating the late time rate of change of complexity without having to take into account the contributions from joint and null boundary terms. In a later work, Myers et al  \cite{Lehner-2016} derive the expressions for the joint and null boundary terms and showed how to include the corresponding contributions to the rate of change of holographic complexity. Myers et al find a perfect match with the results of Brown et al at later times and carefully explain the reasons behind the agreement in \cite{Lehner-2016}. In this appendix we briefly review the approach of Brown et al and we show that it gives the same results obtained in section \ref{sec-HC} using the approach of Myers et al \cite{Lehner-2016,Carmi-2016}.

\begin{figure}
\centering
\includegraphics[width=8.5cm]{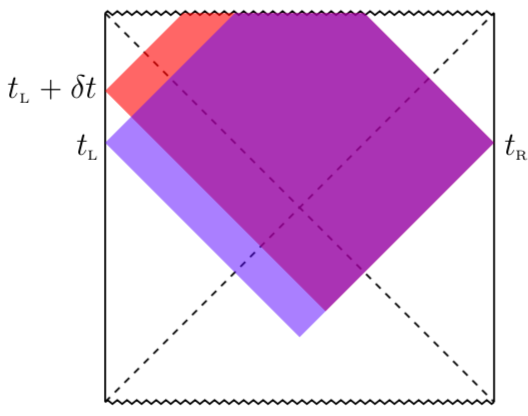}   
\caption{  Change in the WDW patch as the time evolves in the left boundary}
\label{fig-WDW21}
\end{figure}

In the approach of Brown et al it is more convenient to consider the time evolution of the WDW patch when we increase the time in the left boundary, while keeping fixed the time in the right boundary, as shown in figure \ref{fig-WDW21}. 

\begin{figure}
\centering
 \includegraphics[width=6.5cm]{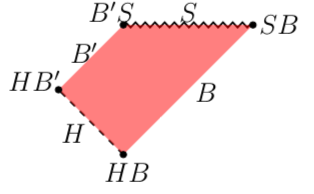}  
\caption{  Piece of the WDW patch that contributes to the rate of change of complexity at late times.}
\label{fig-WDW22}
\end{figure}
 Figure \ref{fig-WDW21} shows that, as the time evolves in the left boundary, the WDW patch increases in the region shown in red, while it decreases in the region shown in light-blue. To calculate the corresponding variation of the WDW patch, the authors of \cite{CA-2} argue as follows:

\begin{itemize}
\item the parts of the WDW patch that lie outside of the horizon are time-independent because this region has a time-translation symmetry. As a consequence, these parts do not contribute to the rate of change of complexity;

\item the part of the WDW patch that lies inside the past horizon contributes at early times, but it is highly suppressed at later times. Hence, at later times, the only contribution for the rate of change of complexity comes from the region of the WDW patch that lies inside the future horizon. This region is shown in  figure \ref{fig-WDW22};

\item under time evolution the surface $B$ is replaced by the surface $B'$, while the corners $HB$ and $SB$ are replaced by the corners $HB'$ and $S'B$, respectively. The surfaces $B$ and $B'$ are related by a time-translation symmetry and so their contributions cancel. The same cancellation occurs between the contributions coming from $HB$ and $HB'$ and between the contributions coming from $SB$ and $B'S$;

\end{itemize}

With the above cancellations the only terms left to be computed are the bulk contribution and the surface contributions coming from the horizon $H$ and from the singularity $S$. Therefore, the gravitational action evaluated in the WDW patch can be written as
\be
I_\mt{WDW}=I_\mt{bulk}+I_\mt{surface}\,,
\ee
where the bulk contribution reads
\be
I_\mt{bulk}=\frac{1}{16\pi G_\mt{N}}  \int_{\mathcal{M}}  d^{d+1}x\sqrt{-g} \mathcal{L}(x)
\ee
while the GHY surface contribution reads
\be
I_{\textrm{surface}}=\frac{1}{8\pi G_\mt{N}}\Big[\int_{r=\rh}d^{d}x\sqrt{|h|}K+\int_{r=\epsilon_0}d^{d}x\sqrt{|h|}K\Big]
\ee
where $r=\rh$ indicates the boundary surface at the horizon and $r=\epsilon_0$ indicates the boundary surface at the singularity. 
For the general action and metric given in (\ref{eq-Sgen}) and (\ref{eq-metricGen}) we can write
\bea
\nonumber I_\mt{WDW}&=&\frac{V_{d-1}}{16\pi G_\mt{N}} \Bigg[  \int dt \int_{\epsilon_0}^{\rh}dr\,\sqrt{-g}\, \mathcal{L}(r)
 \\
&+& \int dt \, \sqrt{\frac{G_{tt}G}{G_{rr}}}\left( \frac{G_{tt}'}{G_{tt}}+\frac{G'}{G} \right)\Big|_{r=\rh} \nonumber \\
&+& \int dt  \sqrt{\frac{G_{tt}G}{G_{rr}}}\left(\frac{G_{tt}'}{G_{tt}}+\frac{G'}{G} \right)\Big|_{r=\epsilon_{0}} \Bigg] \,, 
\eea
where we have used (\ref{eq-K}) to express $K$ in terms of the metric functions. The time-derivative reads
\begin{eqnarray}
\nonumber \frac{d I_\mt{WDW}}{dt}&=&\frac{V_{d-1}}{16\pi G_\mt{N}}  \Big[ \int_{\epsilon_0}^{\rh} \,dr\,\sqrt{-g}\, \mathcal{L}(r)\\
&+& 
 \sqrt{\frac{G}{G_{tt}G_{rr}}} G_{tt}' \Big|_{r=\rh} 
 +\sqrt{\frac{G_{tt}G}{G_{rr}}}\left(\frac{G_{tt}'}{G_{tt}}+\frac{G'}{G} \right)\Big|_{r=\epsilon_{0}}\,,\nonumber \\
\label{CA-suss}
\end{eqnarray}
where we have used that $G_{tt}/G_{rr}$ vanishes at the horizon to simplify the expression for the GHY term at the horizon.
The above results for the late-time rate of change of $I_\mt{WDW}$ precisely coincides with the result (\ref{eq-Iwdw-tlarge}) obtained with the approach of Myers et al \cite{Lehner-2016,Carmi-2016}. The reason for the agreement is the following: both approaches contain identical bulk contributions and identical surface contributions coming from the future singularity. The only difference is that in the calculation of Brown et al there is a GHY-like term for the horizon, while in the calculation of Myers et al there is no such term, but there is instead a joint contribution coming from a corner that lies just behind the past horizon. Surprisingly, these two terms precisely coincides and both approaches give the same result. A more detailed explanation for the agreement between the two approaches can be found in \cite{Lehner-2016}.

\end{document}